\documentclass[10pt,aps,pre,twocolumn,superscriptaddress,nobibnotes,nodoi,reprint,longbibliography]{revtex4-2}

\usepackage{graphicx}
\usepackage{amsmath,physics,bm,float}
\usepackage{xcolor}
\usepackage{soul}
\usepackage[colorlinks=true,citecolor=blue,urlcolor=blue,linkcolor=blue]{hyperref}
\usepackage{amssymb}
\usepackage{bbm}
\usepackage{natbib}
\usepackage{upgreek}
\usepackage{mathtools}
\usepackage{siunitx}
\usepackage[normalem]{ulem}

\def\dd{{\rm d}}

\newcommand{\del}[0]{\partial}
\newcommand{\absgrad}[1]{|\nabla #1|}

\DeclareGraphicsExtensions{.pdf,.png,.jpg}
\graphicspath{{./}}

\begin{document}
\title{Critical escape dynamics of an active particle moving on curved surfaces}

    \author{Maxim Root}
	\affiliation{
		Institut f{\"u}r Theoretische Physik II: Weiche Materie,
		Heinrich-Heine-Universit{\"a}t D{\"u}sseldorf, Universit{\"a}tsstra{\ss}e 1,
		D-40225 D{\"u}sseldorf, 
		Germany}

	\author{Hartmut L{\"o}wen}
	\affiliation{
		Institut f{\"u}r Theoretische Physik II: Weiche Materie,
		Heinrich-Heine-Universit{\"a}t D{\"u}sseldorf, Universit{\"a}tsstra{\ss}e 1,
		D-40225 D{\"u}sseldorf, 
		Germany}

    \author{Peter Sollich}
	\affiliation{
		Institute for Theoretical Physics, University of G\"ottingen, D-37077 G\"ottingen, Germany, 
		Germany}
        \affiliation{
		Department of Mathematics, King’s College London, Strand, London WC2R 2LS, United Kingdom}

    \author{Lorenzo Caprini}
	\affiliation{
		Physics Department, University of Rome La Sapienza, P.le Aldo Moro 5, IT-00185 Rome, Italy}

\author{Alexander P.\ Antonov}
	\email{alexander.antonov@hhu.de}
	\affiliation{
		Institut f{\"u}r Theoretische Physik II: Weiche Materie,
		Heinrich-Heine-Universit{\"a}t D{\"u}sseldorf, Universit{\"a}tsstra{\ss}e 1,
		D-40225 D{\"u}sseldorf, 
		Germany}

        \date{\today}
\begin{abstract}
On the macroscopic scale, active (self-propelled) objects moving through complex environments are often subject to effects arising from both the curvature of the surrounding surface and external influences such as gravity. We study the Kramers escape problem for active Brownian particles in two settings: motion in an external potential driven solely by gradient forces, and motion on a curved manifold subject to an additional constant gravitational force. In the absence of translational noise, a sharp dynamical transition occurs when the self-propulsion velocity $v_0$ reaches a critical velocity $v_c$ required for an escape. Above this threshold, the escape rate $k$ follows a universal scaling form, $k \propto \exp\left[-\mathrm{const}(v_0-v_c)^{-\gamma}\right]$, with an exponent $\gamma=3/2$ for the escape in a potential and $\gamma=1/2$ for the escape on a curved manifold in the presence of gravity. These distinct exponents reveal how the interplay of activity, surface curvature, and gravity determines the critical escape dynamics. Our theoretical predictions are verified by numerical simulations.
\end{abstract}

\maketitle

\section{Introduction}

The escape problem for a particle in an external potential can be traced back to the pioneering work of Kramers \cite{kramers1940brownian}, who studied the escape rate of a Brownian particle from a deep potential well over an energy barrier. In this scenario, escape is driven by thermal fluctuations but remains a rare event because the fluctuations must provide sufficient energy to overcome the potential barrier. As a consequence, the escape rate decreases exponentially with the ratio of the barrier height to the thermal energy, as described by the Arrhenius law \cite{arrhenius1889reaktionsgeschwindigkeit}. Since Kramers' original work, the problem has been extended and applied in numerous contexts, most notably to multidimensional systems \cite{landauer1961frequency}, quantum tunneling problems \cite{larkin1983nonlinear, caldeira1981quantum}, and hydrodynamic flows \cite{yang2017hydrodynamic}; for more details, see the reviews \cite{Hanggi_RMP, melnikov_fifty_years_kramers}.

However, escape events can also be triggered by mechanisms other than thermal activation. A prominent example is \textit{self-propulsion}, or activity, which refers to the ability of an object to convert energy from its environment or internal energy sources into directed motion \cite{marchetti2013hydrodynamics, Bechinger/etal:2016, levis2017active, te2026colloquium,Elgeti/etal:2015}. Self-propulsion enables active particles to overcome potential barriers at rates typically much higher than those of purely Brownian particles, as was first highlighted in early studies of active escape dynamics and barrier crossing \cite{Burada/Lindner:2012,geiseler2016kramers}. Experimental support was later provided in active colloidal systems, where self-propelled particles were observed to cross confining barriers more efficiently than passive ones \cite{Koumakis2016,experiment_Militaru2021}.  
On the theoretical side, several works showed that the escape rate depends sensitively on persistence and on the nonequilibrium statistics of the active force \cite{caprini2019active,Woillez/etal:2019}. Similar acceleration of the escape process has also been explored for non-Gaussian nonequilibrium fluctuations, which, like active forces, drive the system out of equilibrium \cite{baule2023exponential}. Closely related studies also emphasized the role of non-local effects and alternative formulations of active barrier crossing \cite{woillez_nonlocal:2020,Crisanti/Paoluzzi_exactHam:2026}.  
The escape problem has been examined in specific geometries and model systems, including disks \cite{Olsen_disk2020}, harmonic traps \cite{harmonic_Wexler2020}, and other confined settings \cite{Zanovello/etal:2021}. It was also demonstrated that internal degrees of freedom can substantially alter escape pathways and rates \cite{collective_Aranson2022, Tasinkevych/etal:2026}. Recent studies have also addressed active Brownian particles and run-and-tumble particles in this context \cite{Lin_ABP2025,RTP_Basu2026,Huyak/etal:2026,Wei2026,Gueneau/Majumdar:2025}. By contrast, for passive Brownian particles, barrier crossing is governed by standard thermally activated escape and is typically much slower under comparable conditions \cite{Sharma/etal:2017,Chaki2020}.

Remarkably, escape can occur even for athermal particles, i.e., in the absence of translational noise \cite{caprini2021correlated}. In this case, an active particle can overcome a potential barrier only via self-propulsion. Unlike in the passive case, however, the escape rate is not determined solely by the height of the potential barrier. Instead, it depends on the detailed shape of the potential, as the escape is governed by a force balance between self-propulsion and the force exerted by the potential.

Such interplay becomes particularly relevant in realistic environments, in which active particles navigate through complex geometries where external forces and curvature act together \cite{Castro/Sevilla:2018,castro2023,Iyer/etal:2023,Mackay/etal:2026,Caprini/Marconi:2018,Sandoval2018,Webb/etal_MIPS_curved:2026}.\ Of particular interest are curved two‑dimensional manifolds with saddle‑like geometry. On such surfaces, curvature modifies the projection of a constant external force onto the tangent plane, thereby influencing the local direction of motion. Consequently, escape cannot be described as the crossing of a scalar entropic barrier; rather, it emerges from the interplay among self‑propulsion, the projected force field, and the geometric transport dictated by the surface \cite{caprini2019active, Woillez/etal:2019,Mackay/etal:2026,Naji/Brown2007,Ohta/Komura2020,Fily/Baskaran:2016,Li_polar_active:2024,Sknepnek/Henkes:2015}. This leads to a geometry‑controlled dynamical transition that extends the concept of barrier crossing to active matter moving on curved surfaces.

\begin{figure}[t]
    \centering
\includegraphics[width=.7\columnwidth]{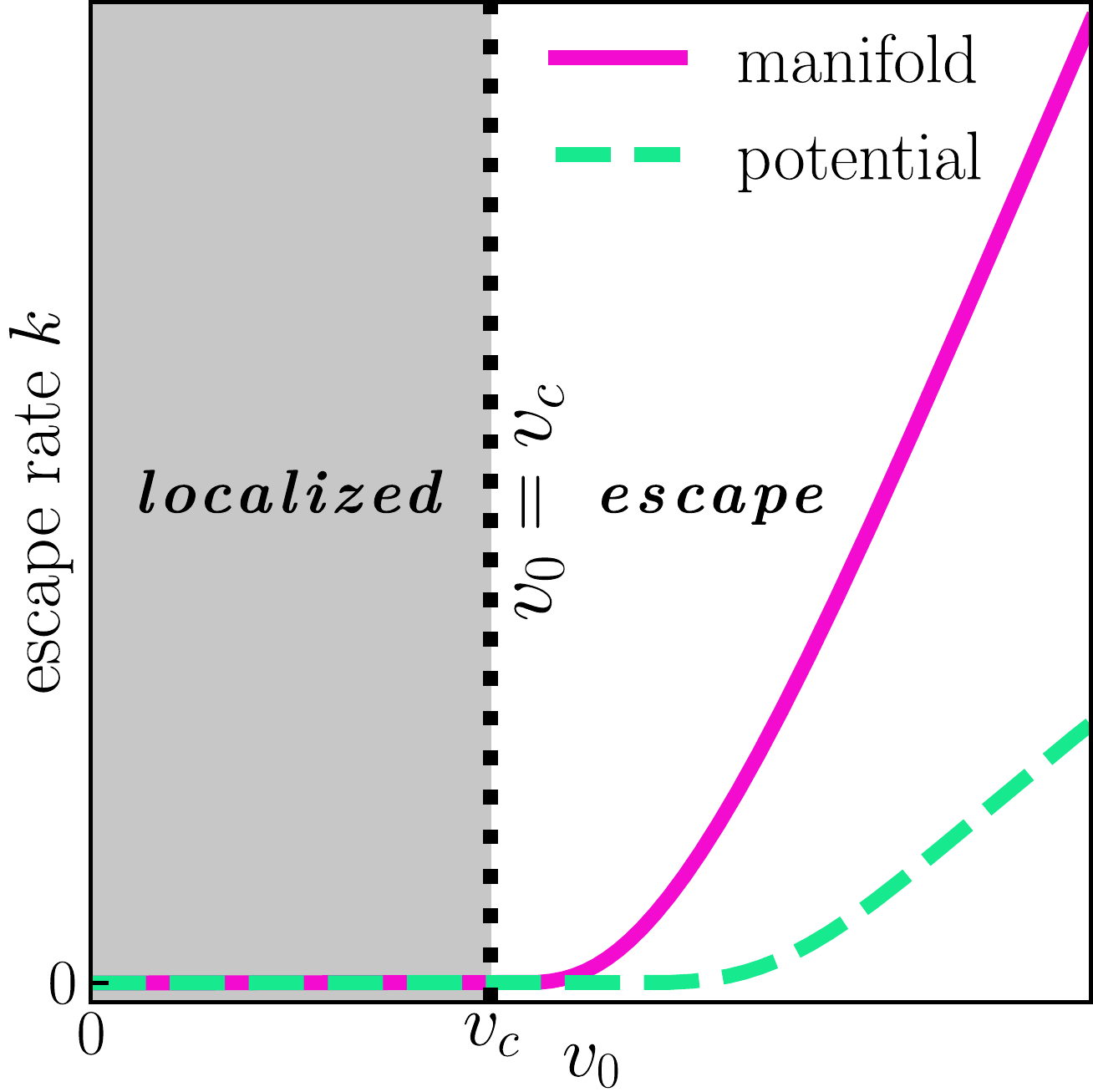}
    \caption{Sketch of the escape rate $k$ for the motion of the active particle in an external potential (green dashed line) and within a curved manifold with additional gravity (magenta solid line). There is a dynamical transition at $v_0=v_c$ (black dotted line) between a localized state ($k=0$) and an escape situation ($k>0$). In the latter regime, the escape rate follows the universal scaling form for $v_0>v_c$ (Eq.~\eqref{eq:scaling}) with $\gamma=3/2$ for the potential and $\gamma=1/2$ for the manifold.}
    \label{fig:1}
\end{figure}

Here, we study the escape dynamics of an active Brownian particle with only rotational noise in two related settings: a saddle potential, and a saddle-shaped manifold subjected to gravity set by a constant external force field. Without the translational noise, escape is only possible when the self-propulsion velocity $v_0$ is above a critical value, $v_c$, set by the maximal opposing force required for an escape. This means, the escape rate $k$ is zero for $v_0\leq v_c$, as shown in Fig.~\ref{fig:1} by the grey region. 
Just above the critical velocity $v_c$, we demonstrate that the escape rate $k$ follows a universal scaling form,
\begin{equation}
\label{eq:scaling}
    k \propto \exp\left[-\mathcal{C}(v_0-v_c)^{-\gamma}\right],
\end{equation}
where $\mathcal{C}$ is a scaling constant that does not depend on the difference $v_0 - v_c$; and the characteristic exponent $\gamma$ is $\gamma=3/2$ for escape in an external potential and $\gamma=1/2$ for escape on a manifold, as summarized in Fig.~\ref{fig:1} in the white region. In both cases, we derive the escape rate in the critical regime where activity is only marginally sufficient to overcome either the potential gradient, or the projected force on the curved surface. In this regime, $v_0 - v_c \ll v_0$, escape is governed by rare orientational fluctuations that keep the particle's orientation along the escape path over timescales much longer than the intrinsic persistence time of the active particle. 

This paper is organized as follows. In Section~\ref{sec:modelA}, we introduce the Kramers problem for an athermal active Brownian particle escaping from a potential well. We then generalize the problem to a particle constrained to a manifold of an analogous shape and consider the additional effect of gravity (Sec.~\ref{sec:modelB}). In Sec.~\ref{sec:theory} we develop a theory predicting the escape events for these two cases. In Sec.~\ref{sec:results} we present the results of the corresponding numerical simulations, which demonstrate an excellent agreement with our theory. In Sec.~\ref{sec:con}, we draw conclusions and outline directions for future research.

\section{Model}
\label{sec:model}

\subsection{Athermal active Brownian particle in an external potential gradient}
\label{sec:modelA}

\begin{figure}[htp!]
    \centering    
    \includegraphics[width=\columnwidth]{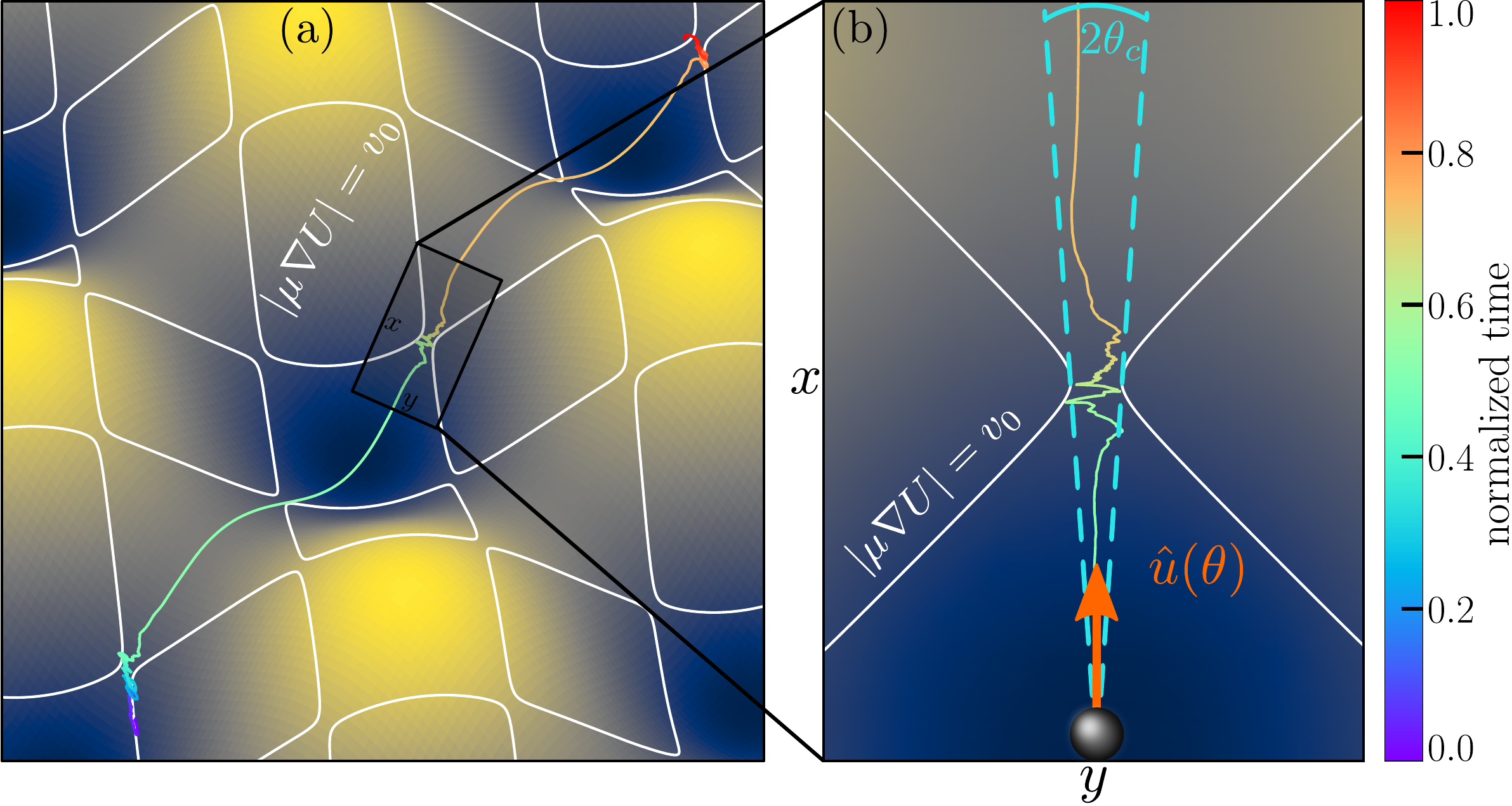}
    \caption{(a) Three-dimensional view of the dynamical motion of an athermal active Brownian particle on the periodic potential $U(x,y)$, with potential maxima colored yellow and minima blue. The white isolines correspond to $|\mu \nabla U| = v_0$, where the active velocity matches the magnitude of the gradient-induced drift; while the particle's position over time is encoded by the trajectory color. (b) Magnified view of the escape bottleneck in the $xy$-plane. Here the particle orientation needs to stay within the orientation interval (lime dashed lines) for a successful escape. The color map on the right corresponds to the time evolution in both panels (a) and (b).}
    \label{fig:2}
\end{figure}

We consider the two-dimensional dynamics of an athermal active Brownian particle (ABP) moving in an external potential $U(\mathbf{r})$, where $\mathbf{r} = (x,y)$ denotes the particle position \cite{Caprini/etal:2021}. The equations of motion are 
\begin{subequations}
\label{eq:EOM-pot}
\begin{align}
\dot{\mathbf{r}} &= v_0 \hat{\mathbf{u}} - \mu \nabla U, \label{eq:translation} \\
\hat{\mathbf{u}} &= (\cos\theta,\sin\theta), \label{eq:rotation} \\
\dot{\theta} &= \sqrt{2D_r}\,\xi(t), \label{eq:noise}
\end{align}
\end{subequations}
where $v_0$ is the bare self-propulsion velocity, $\mu$ represents the mobility, and $D_r$ is the rotational diffusion constant. The stochastic term $\sqrt{2D_r}\,\xi(t)$ accounts for rotational diffusion, with $\xi(t)$ denoting a Gaussian white noise with zero mean $\langle \xi(t) \rangle = 0$ and unit variance $\langle \xi(t)\xi(t') \rangle = \delta(t-t')$. For studying the escape problem, we consider the periodic potential (see the illustration in Fig.~\ref{fig:2}(a))
\begin{equation}
 U(x,y) = \frac{U_0}{2} \left(\sin \left(\frac{2\pi x}{\lambda}\right) - \cos \left(\frac{2\pi y}{\lambda}\right) \right), 
\end{equation}
where $\lambda$ is the potential wavelength and $U_0$ sets the amplitude of the entropic barrier between the neighboring wells. Without loss of generality, we assume that escape occurs along the $x$-direction and that the inflection point is located at the origin, $\mathbf{r} = \mathbf{0}$. In the athermal limit, escape is possible only when the activity is sufficient to overcome the potential force at the inflection point. Since the $x$-directed active motion in Eq.~\eqref{eq:rotation} corresponds to $\theta = 0$, the condition for translational motion in Eq.~\eqref{eq:translation} at the inflection point is
\begin{equation}
    \dot{x} = v_0 - \frac{\pi \mu U_0}{\lambda} \equiv v_0 - v_c > 0,
\end{equation}
which defines the critical velocity, $v_c \equiv \pi \mu U_0/\lambda$, as the maximal force to escape from the potential well.

However, for a successful escape event, the particle must maintain its orientation in close proximity to zero during the whole escape process. More precisely, to hold the $x$-velocity positive in Eq.~\eqref{eq:translation}, the angle must stay within the interval $\theta \in (-\theta_c, \theta_c)$ (see the illustration in Fig.~\ref{fig:2}(b)), where the critical angle $\theta_c$ is determined by the condition
\begin{equation}
    \dot{x} = v_0 \cos\theta_c - v_c = 0.
\end{equation}
In what follows, we consider the escape bottleneck to be narrow, corresponding to small values of the critical angle $\theta_c \ll 1$. In this regime, a second-order Taylor expansion yields
\begin{equation}
\theta_c \approx \sqrt{2\left(1 - \frac{\pi\mu U_0}{\lambda v_0}\right)} = \sqrt{\frac{2(v_0 - v_c)}{v_0}}.
\end{equation}
We introduce the small opening parameter $\delta \ll 1$, 
\begin{equation}
    \delta = \frac{v_0 - v_c}{v_0} = \frac{\theta_c^2}{2},
\end{equation}
to be used as a control parameter governing the escape probability. We consider positive $\delta$ for studies of the escape dynamics, as for $\delta <0$ the particle becomes localized.

\subsection{Geometric contributions from the manifold}
\label{sec:modelB}
\begin{figure}
    \centering
    \includegraphics[width=\linewidth]{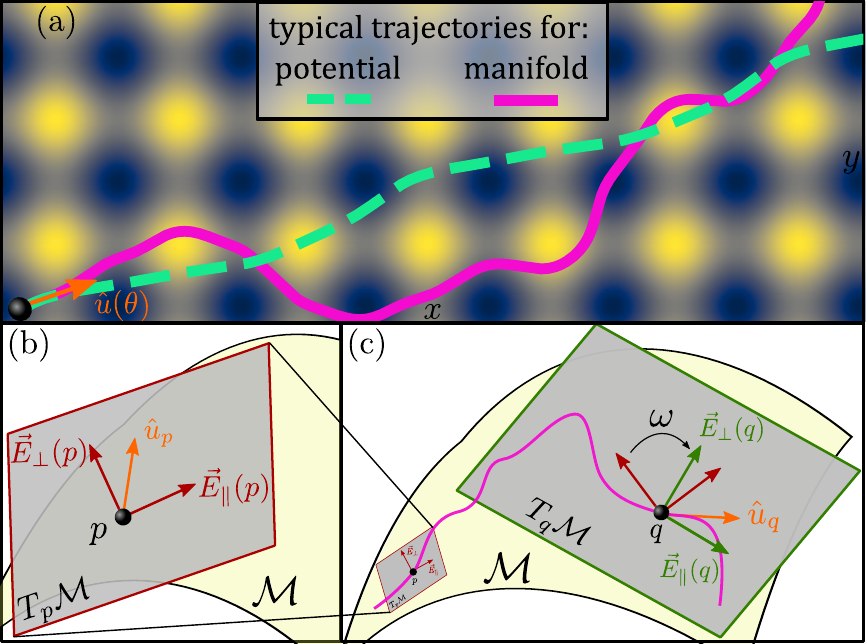}
    \caption{(a) Multiple escape events for a large opening parameter $\delta=3/4$ in the deterministic regime ($D_r=0$), for a particle in the potential landscape (dashed green, Eqs.~\eqref{eq:EOM-pot}) and in the corresponding identical landscape on the manifold with additional gravity (magenta, Eqs.~\eqref{eq:EOM-man}). The initial conditions are identical, $\mathbf{r}(0)=0, \theta = \pi/8$; the difference between the resulting trajectories highlights the impact of geometric contributions from the manifold on the resulting motion. (b) Schematic of an active particle with orientation $\hat u$ moving on a curved manifold $\mathcal{M}$. At point $p\in\mathcal{M}$, the particle orientation is represented in the local orthonormal basis $\{\vec E_\parallel,\vec E_\perp\}$ of the tangent space $T_p\mathcal{M}$. (c) As the particle moves along the curve on the manifold $\mathcal{M}$ to the point $q$, the transport of the local frame between tangent spaces $T_p\mathcal{M}$ and $T_q\mathcal{M}$ induces a rotation of the coordinate axes. This geometric rotation acts as an effective angular velocity and is encoded by the spin connection $\omega$.}
    \label{fig:3}
\end{figure}

Here we start by defining the two-dimensional manifold $\mathcal{M}$ (see Fig.~\ref{fig:3}) and expressing it as an embedding in $\mathbb{R}^3$, following the pedagogical treatment in Ref.~\cite{Nakahara}; for further information, we also refer the reader to Refs.~\cite{Grossmann2015,Nemeth/Adhikari:2025}.
The surface on which the particle is constrained to move is parameterized by $\mathbf{r}(x,y)=(x,y,h(x,y)) \in \mathbb{R}^3$. Here, we take the height profile function $h(x,y)$ to have the same functional form as the potential $U(x,y)$:
\begin{equation}
    h(x,y) = \frac{\lambda}{\pi U_0}U(x,y).
\end{equation}
The metric on the surface thus reads
\begin{subequations}
\begin{equation}
g_{ij}=\mathbf e_i \cdot \mathbf e_j = \delta_{ij}+\del_i h\,\del_j h,
\end{equation}
where the symbol $\cdot$ corresponds to the scalar product while $g_{ij}$ has an inverse metric
\begin{equation}
g^{ij} = \delta^{ij} - \frac{\del^i\!\hspace{.3ex} h\, \del^{\hspace{.1ex}j}\!\hspace{.3ex}h}{g}
\label{eq:metric}
\end{equation}
\end{subequations}
with $g = \det g_{ij}=1+|\nabla h|^2$ corresponding to the matrix determinant. Here co- and contra- variant vectors are related as $\del^i h = g^{ij}\, \del_j h$.
The coordinate basis is given by
\begin{equation}
    \mathbf e_i = \partial_i \mathbf{r} = (\delta_i^{x},\delta_i^{y}, \del_i h),\quad i \in \{x,y\},
\end{equation}
where $\delta_i^j$ are Kronecker delta functions,
\begin{equation}
    \delta_i^j = \begin{cases}
        1, & i=j, \\ 0, & i \ne j.
    \end{cases}
\end{equation}
The homogeneous gravitational force acts in the three-dimensional embedding space and is defined as 
\begin{equation}
    \mathbf F = (0,0,-F_0).
\end{equation}
To obtain the dynamics constrained to the manifold, we project the gravitational force onto the local tangent plane, $F_i =  \mathbf F\cdot \mathbf e_i$. After projection and raising the index, the force becomes proportional to the local height gradient, yielding 
\begin{equation}
F^i= g^{ij} F_j= -\frac{F_0}{g}\del_i h.
\end{equation} 
To describe the self-propulsion of an active particle, we introduce a local orthonormal frame $(\mathbf E_\parallel,\mathbf E_\perp)$, where $\mathbf E_\parallel$ is chosen to align with the surface gradient $\nabla h$. The corresponding frame basis vectors are given by 
\begin{subequations}
\begin{align}
\mathbf E_\parallel &= \frac{1}{\sqrt{g}\absgrad{h}}(\del_x h,\del_y h,\absgrad{h}^2) \\ 
\mathbf E_\perp &= \frac{1}{\absgrad{h}} (-\del_y h,\del_x h,0).
\end{align}
\end{subequations}
The transformation between the coordinate and orthonormal bases is defined through the relation
\begin{equation}
\label{eq:frame_trafo}
\mathbf E_a  = E^i_a \mathbf e_i,\quad a \in \{\parallel,\perp\},
\end{equation}
with the inverse relation obtained analogously.
In the absence of torques, the orientation is parallel transported along the trajectory. As a consequence, the translational dynamics couple to the particle orientation through the rotation of the local frame itself. In two dimensions, this geometric rotation is fully characterized by the spin connection $\omega_i = \mathbf E_\perp \cdot \del_i \mathbf E_\parallel$.
The equations of motion in the coordinate basis then read \cite{Nemeth/Adhikari:2025,Mackay/etal:2026}:
\begin{subequations} 
\label{eq:EOM-man}
\begin{align} 
\dot{r}^i &= v_0 \hat{u}^i + \mu F^i, \label{eq:transl-manifold}\\
\hat{u}^i &= \cos\theta\,E_\parallel^i + \sin\theta\,E_\perp^i, \\ 
\dot{\theta} &= -\omega_i \dot r^i + \sqrt{2D_r}\,\xi(t). \label{eq:coupling}
\end{align}
\end{subequations}
In this scenario, although the basic idea of an escape event illustrated in Fig.~\ref{fig:2}(a) remains the same, the mechanism underlying the escape is modified by the coupling of the particle orientation to the translational dynamics in Eq.~\eqref{eq:coupling}. This is revealed by the difference between the resulting trajectories from Eq.~\eqref{eq:EOM-pot} and Eq.~\eqref{eq:EOM-man} in the deterministic case ($D_r=0$), starting from the same initial conditions (Fig.~\ref{fig:3}(a)). We note that the torque $-\omega_i \dot{r}^i$ in Eq.~\eqref{eq:coupling} persists even in the absence of gravity, when no external force acts on the translational degrees of freedom, because this term is of geometric origin and does not rely on the presence of external forces. In the next section, we develop a theoretical framework within which we highlight this difference.

\section{Analytical prediction for the mean first passage time}
\label{sec:theory}

\subsection{Escape Rate from the Survival Probability}

\subsubsection{Escape in the potential gradient}

To estimate the characteristic timescale of the escape process in the potential gradient, we consider deterministic translational motion at an optimal angle $\theta=0$ along the $x$-axis. Near the inflection point, where the particle reaches its slowest speed during the escape, we approximate the equation of motion \eqref{eq:translation} as
\begin{equation}
    \dot x = v_0\delta + Ax^2
\end{equation}
with $A=2\pi^3 \mu U_0/\lambda^3$.
For the initial condition $x(0)=0$, the solution is
\begin{equation}
x(t) = \sqrt{\frac{v_0\delta}{A}} \tan(\sqrt{Av_0\delta}t).
\end{equation}
Escape requires $x(t)$ to become $O(1)$, which in light of the small prefactor requires the tan to diverge, such that $\sqrt{Av_0\delta}t=\pi/2$. This identifies the characteristic escape timescale:
\begin{equation}
\label{eq:t_e}
t_{\rm e} = \frac{\pi}{2}\frac{1}{\sqrt{Av_0\delta}}.
\end{equation}

The survival problem is formulated in terms of the rotational probability density $p(\theta,t)$ which evolves according to the Fokker-Planck equation
\begin{equation}
\partial_t p = D_r\partial_\theta^2 p,
\label{eq:fpe}
\end{equation}
with the initial condition
\begin{equation}
p(\theta,0)=\delta(\theta).
\label{eq:ic}
\end{equation}
Since survival requires the orientation to remain within the interval $|\theta|<\theta_c$, reaching either of the critical angles $\theta=\pm\theta_c$ terminates the survival process. This is implemented through the absorbing boundary conditions
\begin{equation}
\qquad p(\pm\theta_c,t)=0.
\label{eq:absorb}
\end{equation}
The persistence (or survival) function $P(t)$ is obtained by integrating the probability density over the interval of allowed angles,
\begin{equation}
P(t) = \int_{-\theta_c}^{\theta_c} \dd\theta\, p(\theta,t).
\end{equation}

The general solution can be expressed as an expansion in the eigenmodes of the diffusion operator \cite{Risken1988},
\begin{equation}
p(\theta,t) = \sum_i \phi_i(\theta)e^{-\Lambda_i t},
\end{equation}
where $\phi_i(\theta)$ denotes the eigenfunction associated with the eigenvalue $\Lambda_i$ of the operator $D_r\partial_\theta^2$, subject to the absorbing boundary conditions in Eq.~\eqref{eq:absorb}. The eigenvalues are ordered as
\begin{equation}
0<\Lambda_1<\Lambda_2<\ldots,
\end{equation}
and at long times, the mode with the smallest non-zero eigenvalue $\Lambda_1$ governs the diffusive process. Therefore, the persistence function decays exponentially as
\begin{equation}
P(t)\propto e^{-\Lambda_1 t}.
\end{equation}
$\Lambda_1$ can be interpreted as the \textit{angular decay rate} for remaining within the angular interval required for escape. 
The absorbing boundary conditions determine its dependence on the critical angle, yielding
\begin{equation}
\Lambda_1 = \frac{\pi^2}{4}\frac{D_r}{\theta_c^2},
\label{eq:angular_survival_rate}
\end{equation}
such that the expression for $P(t)$ reads:
\begin{equation}
    P(t) \propto e^{-\Lambda_1 t} = e^{-\frac{\pi^2 D_rt}{4\theta_c^2}}.
\end{equation}
Thus, approaching the critical point suppresses escape in two ways: the characteristic time $t_{\rm e}$ increases, while simultaneously the probability of maintaining a favorable orientation over this time decreases because the allowed angular interval shrinks.
At the characteristic escape timescale, $t=t_{\rm e}$, the persistence function reads
\begin{equation}
    P_{\rm e} \equiv P(t_{\rm e}) \propto \exp\left(-\frac{\pi^3 D_r}{16\sqrt{A v_0}\,\delta^{3/2}}\right)
    \label{eq:Pe-potential}
\end{equation}
with $\delta = (v_0-v_c)/v_0$, and can be interpreted as the escape probability, conditioned on the particle being oriented towards the escape direction. This yields the scaling form announced in Eq.~\eqref{eq:scaling} and the link to the escape rate $k$ will be further discussed in Sec.~\ref{sec:theory}.

\subsubsection{Escape on the manifold}
To obtain insight into the dynamics on the manifold, we first derive the explicit form of the spin connection components, followed by the simplification of the resulting equations of motion using the Taylor expansion near the inflection point.

The transformation of the force into the gradient-aligned basis introduced in Sec.~\ref{sec:modelB} gives 

\begin{subequations}
\begin{align}
    F^\parallel &= -F_0 \frac{\absgrad{h}}{\sqrt{g}}, \\ F^\perp &= 0.  
\end{align}
\end{subequations}
Since the orientation $u^i$ is already defined in the gradient-aligned frame, the spatial drift components of \eqref{eq:transl-manifold} take the form 
\begin{subequations}
\begin{align}
    \dot x^\parallel &= v_0 \cos \theta - \mu F_0 \frac{\absgrad{h}}{\sqrt{g}}, \\ 
    \dot x^\perp &= v_0 \sin \theta.
\end{align}
\end{subequations}
Similar to escape in a potential gradient, one can see that along the escape direction $\theta=0$, the active particle can only overcome the gradient if 
\begin{equation}
v_0-\mu F_0 \max_{y=0} \left[ \frac{\absgrad{h}}{\sqrt{g}}\right] \equiv v_0 - v_c>0 .
\end{equation}
Just as in the potential case, $\absgrad{h}$ attains its maximum at the inflection point at the origin, where the surface is steepest and, consequently, the downhill force acting on the particle is strongest. This gives $\max\absgrad{h}=1$ and since $g=1+\absgrad{h}^2$, the critical velocity reads:
\begin{equation}
    v_c = \frac{\mu F_0}{\sqrt2}.
\end{equation}

The spin connection components are defined as
\begin{align}
\omega_i
= \mathbf E_\perp \cdot \partial_i \mathbf E_\parallel
= \frac{\del_x h\, \del^2_{yi}h-\del_y h\, \del^2_{xi}h}{\sqrt{g}|\nabla h|^2}\, .
\end{align}
In the gradient-aligned basis, the components of the spin connection take the form
\begin{subequations}
\begin{align}
\omega_\parallel
&= \frac{\del_x h\del_y h}{g|\nabla h|^3}(\del^2_{yy}h-\del^2_{xx}h),\\
\omega_\perp
&= \frac{1}{\sqrt{g}|\nabla h|^3}
\left((\del_x h)^2\del^2_{yy}h+(\del_y h)^2\del^2_{xx}h\right),
\end{align}
\end{subequations}
where we have used $\del^2_{xy}h=\del^2_{yx}h=0$ for the surface $h(x,y)$ considered here to reduce clutter.
The rotational drift is obtained by contracting the spin connection with the particle velocity \cite{Mackay/etal:2026,Nemeth/Adhikari:2025},
\begin{align}
\label{eq:spin}
\omega_a\dot{x}^a&
=
\frac{\del_x h\del_y h}{g|\nabla h|^3}(\del^2_{yy}h-\del^2_{xx}h)
\left(
v_0\cos\theta
-\mu F_0\frac{|\nabla h|}{\sqrt g}
\right)\nonumber\\
&+
\frac{(\del_x h)^2\del^2_{yy}h+(\del_y h)^2\del^2_{xx}h}
{\sqrt g|\nabla h|^3}
v_0\sin\theta.
\end{align}
To obtain the behavior of the rotational drift, $\omega_a \dot x^a$, around the inflection point $x=y=0$, we first Taylor expand $h(x,y)$ to obtain
\begin{equation}
h(x,y)-h_0=x+\frac{\pi}{\lambda}y^2-\frac{2\pi^2}{3\lambda^2}x^3 +\mathcal{O}(x^5,y^4)
\end{equation}
with $h_0=-\lambda/(2\pi)$. Inserting this into Eq.~\eqref{eq:spin} and expanding to linear order in $x$, $y$ and $\theta$ gives
\begin{equation}
\label{eq:spin_lin}
\omega_a\dot{x}^a
\approx \alpha\theta+\beta y
\end{equation}
with
\begin{equation}
\alpha
=v_0\frac{\sqrt{2}\pi}{\lambda},
\qquad
\beta=v_0 \delta
\frac{2\pi^2}{\lambda^2}.
\end{equation}
In the considered regime of small $\delta$, the coefficient $\beta$ is parametrically small, whereas $\alpha$ remains finite. Consequently, $\alpha\gg\beta$, and the dominant contribution to the geometric rotational drift is the term $\alpha\theta$.

To derive similarly the leading order terms in the equations of motion for $x$ and $y$, note that sufficiently close to the inflection point, the gradient-aligned frame is approximately Cartesian and given by
\begin{subequations}
\begin{align}
    \mathbf E_\parallel &\approx \frac{1}{\sqrt 2}(1,0,1), \\
    \mathbf E_\perp &\approx(0,1,0).
\end{align}
\end{subequations}
Taylor expanding around the inflection point and for small $\theta$ then gives to leading order
\begin{subequations}
\label{eq:linear}
\begin{align}
\dot x &\approx \frac{v_0}{\sqrt2}\left( \delta-\frac{\theta^2}{2} \right)+\tilde{A}(x^2 - y^2),\\
\dot y &\approx v_0\, \theta
\end{align}
with $\tilde{A} = \sqrt2\pi^2 \mu F_0/\lambda^2$, while for the angle, adding the stochastic component to Eq.~\eqref{eq:spin_lin},
\begin{equation}
\dot \theta \approx -\alpha\,\theta +\sqrt{2D_r}\,\xi(t) .
\label{eq:linear_c}
\end{equation}
\end{subequations}
Note that in Eq.~\eqref{eq:linear_c} the rotation is no longer coupled to the translational motion -- the only geometric contribution is a restoring torque $\alpha\, \theta$. 

However, the presence of this restoring torque qualitatively changes the escape rate. The previous argument for the survival probability no longer applies because, in the presence of the restoring torque, the particle may briefly escape the interval $[-\theta_c, \theta_c]$ and subsequently return due to the confining torque. Such excursions, therefore, do not necessarily interrupt the escape process. This is not the case for free angular diffusion. There, brief excursions are much rarer, and the particle tends to undergo longer excursions outside the interval $[-\theta_c, \theta_c]$, which result in failed escape attempts. Therefore, to estimate the escape rate, we consider the survival probability for the full Fokker-Planck dynamics for $p(x,y,\theta;t)$, corresponding to the Langevin equations \eqref{eq:linear}:
\begin{align}
\label{eq:FPE}
\del_t p
= &-\partial_x\!\left[\left(\tilde{A}(x^2-y^2)+\frac{v_0}{\sqrt2}\left(\delta-\frac{\theta^2}{2}\right)\right)p\right]\nonumber\\
&- v_0\theta\,\partial_y p
+ \alpha\,\del_\theta(\theta p) + D_r\,\del_\theta^2 p\,.
\end{align}
We note that, despite the presence of a confining torque, escape in the manifold case does not become quasideterministic (i.e., does not occur on the first attempt), even in the small-fluctuation limit, $\langle \theta^2\rangle = D_r/\alpha \ll \delta$. This is because the force in the $x$-direction in Eq.~\eqref{eq:FPE} additionally contains a $-y^2$ term that prevents the free escape. Its average, for $y(0)=0$, is given by \cite{Bechinger/etal:2016}
\begin{equation}
\langle y^2 \rangle = \frac{2 v_0^2 D_r}{\alpha^2}
\left(t-\frac{1-e^{-\alpha t}}{\alpha}\right),
\end{equation}
which grows quadratically at short times, $t\ll\alpha^{-1}$, and diffusively at long times, $t\gg\alpha^{-1}$.

We discuss the procedure for estimating the escape rate in the following subsection.

\subsection{Mean first passage time}

The overall escape mechanism for the potential gradient can be described as follows. First, the particle must, through stochastic fluctuations, become oriented towards the escape path. It must then maintain the correct orientation for a sufficiently long time to overcome the inflection point, which occurs with probability $P_{\rm e}$ derived in the previous subsections. Consequently, the number of failed attempts $n_f$ before the first successful attempt follows a geometric distribution $(1-P_{\rm e})^{n_f}P_{\rm e}$, assuming that successive attempts are independent and have the same success probability \cite{vanKampen}.

The corresponding mean number of failed attempts is
\begin{equation}
\langle n_f \rangle = \frac{1-P_{\rm e}}{P_{\rm e}}.
\end{equation}
This gives the mean first-passage time (MFPT) 
\begin{equation}
\langle \tau \rangle
= t_{\rm e}
+ \frac{1-P_{\rm e}}{P_{\rm e}}
\left(
\langle t_f\rangle + \langle t_w\rangle
\right),
\end{equation}
where $\langle t_f\rangle$ and $\langle t_w\rangle$ denote the mean durations of a failed attempt and the orienting (waiting) period, respectively.
Since the (un)successful escape times $t_{\rm e}$ ($t_{f}$) are negligible compared with the total waiting time, we obtain
\begin{equation}
\langle \tau\rangle
\simeq \frac{\langle t_w\rangle}{P_{\rm e}}.
\label{eq:MFPT}
\end{equation}
The inverse, $1/\langle\tau\rangle$, corresponds to the intuitive ``attempt frequency $\times$ success probability'' picture, where $1/\langle t_w\rangle$ is the attempt frequency and $P_{\rm e}$ is the probability of a successful escape per attempt.

In the rare-escape limit, the typical number of failed attempts before escape is large,
\begin{equation}
n_f = \frac{t}{\langle t_w\rangle}\gg1.
\end{equation}
In this continuous-time limit, the geometric distribution approaches an exponential distribution. Using the relation
\begin{equation}
(1-x)^m \simeq \exp(-mx),
\qquad x\ll1,
\end{equation}
the probability density for escape at time $\tau = t$ becomes
\begin{align}
p(\tau=t)
&\simeq
\frac{P_{\rm e}}{\langle t_w\rangle}
\left(1-P_{\rm e}\right)^{t/\langle t_w\rangle}
\simeq
\frac{P_{\rm e}}{\langle t_w\rangle}
\exp\left(
-\frac{P_{\rm e}}{\langle t_w\rangle}t
\right).
\end{align}
Thus, the first-passage-time distribution is exponential,
\begin{equation}
p(\tau)
= k e^{-k\tau},
\qquad
k=\frac{P_{\rm e}}{\langle t_w\rangle}
=\frac{1}{\langle\tau\rangle}.
\label{eq:poisson}
\end{equation}
This corresponds to a Poisson escape process with rate $k$ \cite{Ross1996}. 

For the manifold, rather than estimating the survival probability, we calculate the escape rate directly by implementing the absorbing boundary condition for the escape itself, $p(x=\lambda/2,t)=0$. The escape rate is then nothing but the lowest eigenvalue of the Fokker-Planck equation, \eqref{eq:FPE},
\begin{equation}
\label{eq:escape-manifold}
    k \approx \tilde{\varLambda}_1.
\end{equation}
We note that $\tilde{\varLambda}$ and $\Lambda$, although both are eigenvalues of the corresponding Fokker-Planck equations, describe different physical processes: the former characterizes the escape probability, whereas the latter characterizes the survival probability. 

In the Results section, we calculate the eigenvalue $\tilde{\varLambda}_1$ numerically due to the overall complexity of the analysis for the Fokker-Planck equation \eqref{eq:FPE}. We use the standard approach based on the backward Fokker-Planck equation with an absorbing boundary condition \cite{gardiner2009stochastic}. For the numerical calculation, the backward generator is discretized on a uniform three-dimensional grid in $(x,y,\theta)$ using finite differences. The drift terms are discretized using an upwind scheme, while the angular diffusion term is approximated by a standard centered second-order finite difference. The principal eigenvalue is obtained using an implicitly restarted Arnoldi method as implemented in ARPACK \cite{arnoldi1951principle}, targeting the eigenvalue with the largest real part, which corresponds to the eigenvalue closest to zero for the present generator.

\section{Results}
\label{sec:results}

\begin{figure}[t]
    \centering
    \includegraphics[width=\linewidth]{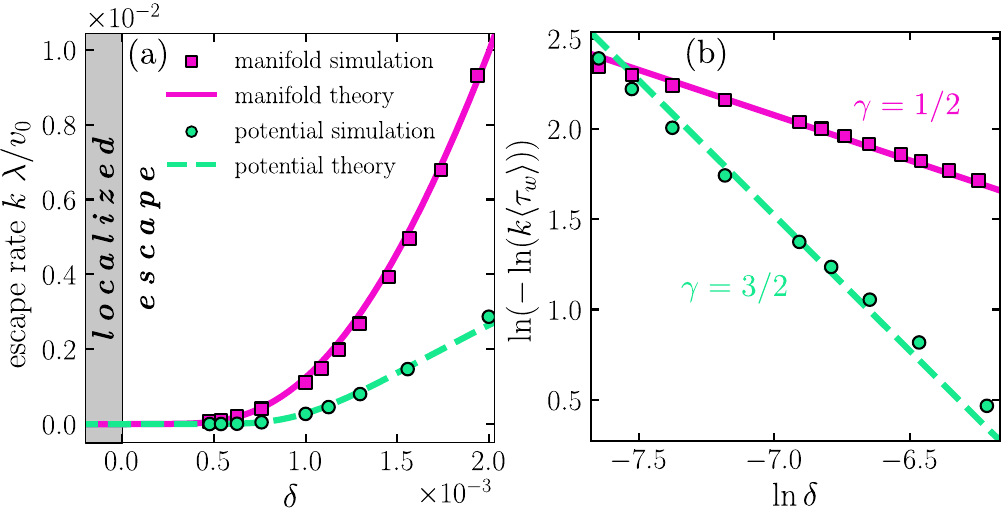}
    \caption{Escape rate $k$ for various opening parameters $\delta$ on (a) a linear scale and (b) as $\ln(-\ln( k \langle t_w \rangle))$ vs $\ln \delta$, where the slope yields 
the exponent $\gamma$ in the scaling \eqref{eq:scaling} $k\langle t_w\rangle \sim \exp(-\delta^{-\gamma})$; plotted for potential gradient (dashed green) and manifold (solid magenta) obtained analytically (Eq.~\eqref{eq:Pe-potential} with Eq.~\eqref{eq:poisson} for potential gradient and Eq.~\eqref{eq:escape-manifold} for manifold). The prefactors of the exponential laws are adjusted to match the numerical data (dots), with the resulting agreement in the scaling trends supporting the analytical predictions. Panel (a) reveals faster escape from the manifold, while panel (b) makes the characteristic exponents of Eq.~\eqref{eq:scaling} apparent. The scaling constant $\mathcal C$ is fitted from the numerical solution of the linearized Fokker–Planck equation \eqref{eq:FPE} to simulation data from the full Langevin dynamics for the manifold, while for the potential gradient, we use the analytical expression $\mathcal{C} = \pi^{3/2} D_r v_0 \lambda^{3/2} / (16\sqrt{2\mu U_0})$ obtained from Eq.~\eqref{eq:Pe-potential}. Parameters used: $D_r=10^{-4} v_0/\lambda$, and $U_0$ and $F_0$ corresponding to the range of $\delta$ considered are given by $F_0=\sqrt{2}(1-\delta)v_0/\mu,\, U_0=\lambda (1-\delta) v_0/(\pi\mu)$.}
    \label{fig:4}
\end{figure}

\begin{figure}[t]
    \centering
    \includegraphics[width=\linewidth]{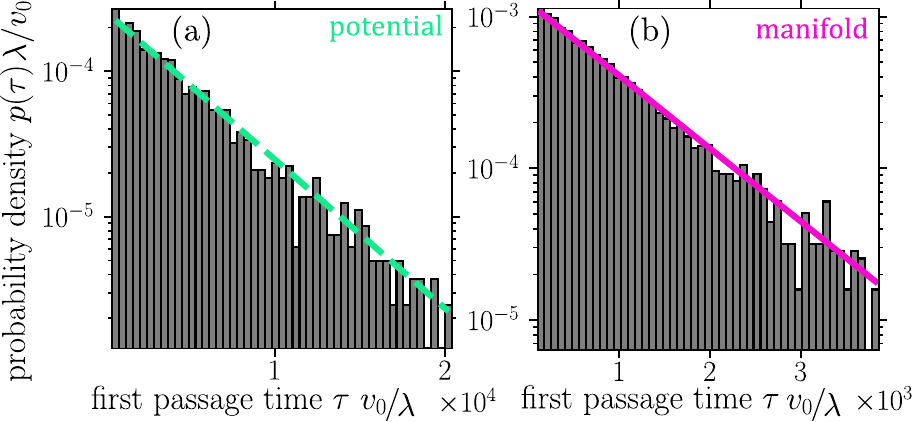}
    \caption{Distribution of escape (first passage) times in semi-log scaling, obtained from numerical simulations of (a) potential gradient, Eq.~\eqref{eq:EOM-pot}, and (b) manifold, Eq.~\eqref{eq:EOM-man}.\ Parameters used: $D_r=10^{-4}v_0/\lambda, \, \delta=10^{-3},\, F_0=\sqrt{2}(1-\delta)v_0/\mu,\, U_0=\lambda (1-\delta) v_0/(\pi\mu)$. The (a) dashed green and (b) solid magenta lines correspond to the exponential distributions with rates (a) $k=2.36\times10^{-4} \ v_0/\lambda$ and (b) $k=1.11\times10^{-3} \ v_0/\lambda$ obtained from simulations and serve as guides for an eye.}
    \label{fig:5}
\end{figure}

In Figure \ref{fig:4}(a), we show that the analytically predicted escape rates are in excellent agreement with the numerical results. To obtain the numerical escape rates, we numerically solve the equations of motion \eqref{eq:EOM-pot}, \eqref{eq:EOM-man} using the Euler-Maruyama scheme for a time step $\Delta t = 10^{-3}\lambda/v_0$, and determine the corresponding mean first passage time $\langle \tau \rangle$, defined as the time required for the particle to reach the vicinity of another potential well. The escape rate is then calculated as $k=1/\langle \tau \rangle$. The resulting distribution of escape times is consistent with a Poisson process, as described by Eq.~\eqref{eq:poisson} and confirmed by the numerical results shown in Fig.~\ref{fig:5}.

Our results confirm that the escape rates follow the universal scaling law suggested in Eq.~\eqref{eq:scaling}.\ However, due to different escape mechanisms, the characteristic exponent changes from $\gamma = 3/2$ in the potential case to $\gamma = 1/2$ in the manifold case (Fig.~\ref{fig:4}(b)). 
This means that the faster transitions in the manifold case cannot be attributed merely to a quantitative change in the prefactor of the exponential law; instead, they represent a genuine exponential increase. By contrast to the potential-gradient mechanism, where successful escape relies on the orientation surviving within the escape window long enough for the particle to cross the barrier, the manifold introduces a restoring torque that actively drives the particle orientation towards the escape direction. This orientational drift plays a decisive role even in the presence of rotational diffusion, thereby substantially enhancing the probability of successful escape. The numerical results also confirm that the escape dynamics on the manifold is effectively captured by the simplified system described by Eqs.~\eqref{eq:linear}.

In fact, if we suggest that the escape rate for the manifold case has the same functional form as the escape rate for the potential,
$k \propto \exp(-\Lambda_1 t_{\rm e}),\, t_{\rm e} \propto 1/\sqrt{\delta}$, given by Eqs.~\eqref{eq:Pe-potential} and \eqref{eq:poisson}, it appears that $\Lambda_1$ is effectively independent of $\delta$ as the correct exponent $\gamma=1/2$ is already present in the characteristic escape time $t_{\rm e}$. This independence can be explained in the following way: for the manifold, the effective restoring torque confines the orientation towards the escape direction and counteracts diffusive excursions outside the angular interval. As a result, $\Lambda_1$ becomes insensitive to the decreasing width of the survival interval as $\delta \to 0$. The dominant $\delta$-dependence of the escape rate is therefore solely generated by the increasing time $t_{\rm e} \propto 1/\sqrt{\delta}$ for which the favorable configuration has to be maintained. This argument naturally leads to $k \propto \exp(-\Lambda_1 t_{\rm e})$ with an effectively $\delta$-independent $\Lambda_1$. In contrast, for the potential case, there is no stabilizing mechanism that opposes free angular diffusion, and it becomes increasingly difficult to survive in the shrinking escape window, as follows from the dependence $\Lambda_1 \propto \delta^{-1}$. This qualitative difference highlights the role of the geometric coupling between translational and rotational motion in the manifold case.

\section{Conclusions and outlook}
\label{sec:con}

In this paper, we have studied the Kramers' problem for active Brownian particles with orientational noise in two paradigmatic systems: in a potential gradient and on a manifold. We demonstrate that, in spite of the similar mechanism behind it, the confinement to the curved geometry exponentially accelerates the escape process, changing the exponent in the universal scaling law \eqref{eq:scaling} from $\gamma=3/2$ to $\gamma=1/2$. Our results have great practical significance in both natural and engineered systems \cite{palagi2018bioinspired} for situations in which motion along physical surfaces is dominated by self-propulsion \cite{Bechinger/etal:2016}. 
Future perspectives include the possibility of introducing intelligence \cite{cichos2020machine, lowen2026towards, jeggle2026intelligent, olsen2026information} to an active particle for efficient navigation in complex environments, and the exploration of Mpemba-like phenomena in multistable potential landscapes \cite{Schwarzendahl/Lowen:2022, Antonov/Lowen:2026}. Additionally, it would be interesting to investigate an optimization problem in which the angular torque arising from the geometry suppresses escape, and to determine whether an optimal rotational diffusion coefficient exists that minimizes the escape time \cite{debnath2021escape, upadhyaya2024narrow}. A prospective experimental platform for implementing and testing our results is active granular matter consisting of vibrating robots \cite{baconnier2022selective, Chor/etal:2023, engbring2023nonlinear, caprini2024emergent, antonov2024inertial, antonov2025self, goerlich2026particle}, which provides a direct route for probing the escape dynamics of active particles subject to gravity and constrained to a tunable saddle-shaped surface.

\section*{Acknowledgments}
 A.A.~acknowledges funding from the Deutsche Forschungsgemeinschaft (DFG, German Research Foundation) -- Project-ID 570812137. H.L.~acknowledges funding from the DFG within project LO 418/29-1. 


\begin{thebibliography}{72}%
	\makeatletter
	\providecommand \@ifxundefined [1]{%
		\@ifx{#1\undefined}
	}%
	\providecommand \@ifnum [1]{%
		\ifnum #1\expandafter \@firstoftwo
		\else \expandafter \@secondoftwo
		\fi
	}%
	\providecommand \@ifx [1]{%
		\ifx #1\expandafter \@firstoftwo
		\else \expandafter \@secondoftwo
		\fi
	}%
	\providecommand \natexlab [1]{#1}%
	\providecommand \enquote  [1]{``#1''}%
	\providecommand \bibnamefont  [1]{#1}%
	\providecommand \bibfnamefont [1]{#1}%
	\providecommand \citenamefont [1]{#1}%
	\providecommand \href@noop [0]{\@secondoftwo}%
	\providecommand \href [0]{\begingroup \@sanitize@url \@href}%
	\providecommand \@href[1]{\@@startlink{#1}\@@href}%
	\providecommand \@@href[1]{\endgroup#1\@@endlink}%
	\providecommand \@sanitize@url [0]{\catcode `\\12\catcode `\$12\catcode
		`\&12\catcode `\#12\catcode `\^12\catcode `\_12\catcode `\%12\relax}%
	\providecommand \@@startlink[1]{}%
	\providecommand \@@endlink[0]{}%
	\providecommand \url  [0]{\begingroup\@sanitize@url \@url }%
	\providecommand \@url [1]{\endgroup\@href {#1}{\urlprefix }}%
	\providecommand \urlprefix  [0]{URL }%
	\providecommand \Eprint [0]{\href }%
	\providecommand \doibase [0]{https://doi.org/}%
	\providecommand \selectlanguage [0]{\@gobble}%
	\providecommand \bibinfo  [0]{\@secondoftwo}%
	\providecommand \bibfield  [0]{\@secondoftwo}%
	\providecommand \translation [1]{[#1]}%
	\providecommand \BibitemOpen [0]{}%
	\providecommand \bibitemStop [0]{}%
	\providecommand \bibitemNoStop [0]{.\EOS\space}%
	\providecommand \EOS [0]{\spacefactor3000\relax}%
	\providecommand \BibitemShut  [1]{\csname bibitem#1\endcsname}%
	\let\auto@bib@innerbib\@empty
	\bibitem [{\citenamefont {Kramers}(1940)}]{kramers1940brownian}%
	\BibitemOpen
	\bibfield  {author} {\bibinfo {author} {\bibfnamefont {H.~A.}\ \bibnamefont
			{Kramers}},\ }\bibfield  {title} {\bibinfo {title} {{B}rownian motion in a
			field of force and the diffusion model of chemical reactions},\ }\href
	{https://doi.org/10.1016/s0031-8914(40)90098-2} {\bibfield  {journal}
		{\bibinfo  {journal} {Physica}\ }\textbf {\bibinfo {volume} {7}},\ \bibinfo
		{pages} {284} (\bibinfo {year} {1940})}\BibitemShut {NoStop}%
	\bibitem [{\citenamefont
		{Arrhenius}(1889)}]{arrhenius1889reaktionsgeschwindigkeit}%
	\BibitemOpen
	\bibfield  {author} {\bibinfo {author} {\bibfnamefont {S.}~\bibnamefont
			{Arrhenius}},\ }\bibfield  {title} {\bibinfo {title} {{{{{{\"U}ber}}}} die
			{{R}eaktionsgeschwindigkeit} bei der {{I}nversion} von {{R}ohrzucker} durch
			{{S}{\"a}uren}},\ }\href {https://doi.org/10.1515/zpch-1889-0408} {\bibfield
		{journal} {\bibinfo  {journal} {Z. Phys. Chem.}\ }\textbf {\bibinfo {volume}
			{4}},\ \bibinfo {pages} {226} (\bibinfo {year} {1889})}\BibitemShut {NoStop}%
	\bibitem [{\citenamefont {Landauer}\ and\ \citenamefont
		{Swanson}(1961)}]{landauer1961frequency}%
	\BibitemOpen
	\bibfield  {author} {\bibinfo {author} {\bibfnamefont {R.}~\bibnamefont
			{Landauer}}\ and\ \bibinfo {author} {\bibfnamefont {J.}~\bibnamefont
			{Swanson}},\ }\bibfield  {title} {\bibinfo {title} {Frequency factors in the
			thermally activated process},\ }\href
	{https://doi.org/10.1103/physrev.121.1668} {\bibfield  {journal} {\bibinfo
			{journal} {Phys. Rev.}\ }\textbf {\bibinfo {volume} {121}},\ \bibinfo {pages}
		{1668} (\bibinfo {year} {1961})}\BibitemShut {NoStop}%
	\bibitem [{\citenamefont {Larkin}\ and\ \citenamefont
		{Ovchinnikov}(1975)}]{larkin1983nonlinear}%
	\BibitemOpen
	\bibfield  {author} {\bibinfo {author} {\bibfnamefont {A.~I.}\ \bibnamefont
			{Larkin}}\ and\ \bibinfo {author} {\bibfnamefont {Y.~N.}\ \bibnamefont
			{Ovchinnikov}},\ }\bibfield  {title} {\bibinfo {title} {Nonlinear
			conductivity of superconductors in the mixed state},\ }\href
	{https://www.jetp.ras.ru/cgi-bin/dn/e_041_05_0960.pdf} {\bibfield  {journal}
		{\bibinfo  {journal} {JETP Letters}\ }\textbf {\bibinfo {volume} {41}},\
		\bibinfo {pages} {960} (\bibinfo {year} {1975})}\BibitemShut {NoStop}%
	\bibitem [{\citenamefont {Caldeira}\ and\ \citenamefont
		{Leggett}(1981)}]{caldeira1981quantum}%
	\BibitemOpen
	\bibfield  {author} {\bibinfo {author} {\bibfnamefont {A.~O.}\ \bibnamefont
			{Caldeira}}\ and\ \bibinfo {author} {\bibfnamefont {A.~J.}\ \bibnamefont
			{Leggett}},\ }\bibfield  {title} {\bibinfo {title} {Influence of dissipation
			on quantum tunneling in macroscopic systems},\ }\href
	{https://doi.org/10.1103/physrevlett.46.211} {\bibfield  {journal} {\bibinfo
			{journal} {Phys. Rev. Lett.}\ }\textbf {\bibinfo {volume} {46}},\ \bibinfo
		{pages} {211} (\bibinfo {year} {1981})}\BibitemShut {NoStop}%
	\bibitem [{\citenamefont {Yang}\ \emph {et~al.}(2017)\citenamefont {Yang},
		\citenamefont {Liu}, \citenamefont {Li}, \citenamefont {Marchesoni},
		\citenamefont {H{\"a}nggi},\ and\ \citenamefont
		{Zhang}}]{yang2017hydrodynamic}%
	\BibitemOpen
	\bibfield  {author} {\bibinfo {author} {\bibfnamefont {X.}~\bibnamefont
			{Yang}}, \bibinfo {author} {\bibfnamefont {C.}~\bibnamefont {Liu}}, \bibinfo
		{author} {\bibfnamefont {Y.}~\bibnamefont {Li}}, \bibinfo {author}
		{\bibfnamefont {F.}~\bibnamefont {Marchesoni}}, \bibinfo {author}
		{\bibfnamefont {P.}~\bibnamefont {H{\"a}nggi}},\ and\ \bibinfo {author}
		{\bibfnamefont {H.}~\bibnamefont {Zhang}},\ }\bibfield  {title} {\bibinfo
		{title} {Hydrodynamic and entropic effects on colloidal diffusion in
			corrugated channels},\ }\href {https://doi.org/10.1073/pnas.1707815114}
	{\bibfield  {journal} {\bibinfo  {journal} {Proc. Nat. Acad. Sci.}\ }\textbf
		{\bibinfo {volume} {114}},\ \bibinfo {pages} {9564} (\bibinfo {year}
		{2017})}\BibitemShut {NoStop}%
	\bibitem [{\citenamefont {H{\"a}nggi}\ \emph {et~al.}(1990)\citenamefont
		{H{\"a}nggi}, \citenamefont {Talkner},\ and\ \citenamefont
		{Borkovec}}]{Hanggi_RMP}%
	\BibitemOpen
	\bibfield  {author} {\bibinfo {author} {\bibfnamefont {P.}~\bibnamefont
			{H{\"a}nggi}}, \bibinfo {author} {\bibfnamefont {P.}~\bibnamefont
			{Talkner}},\ and\ \bibinfo {author} {\bibfnamefont {M.}~\bibnamefont
			{Borkovec}},\ }\bibfield  {title} {\bibinfo {title} {Reaction-rate theory:
			fifty years after {K}ramers},\ }\href
	{https://doi.org/10.1103/revmodphys.62.251} {\bibfield  {journal} {\bibinfo
			{journal} {Rev. Mod. Phys.}\ }\textbf {\bibinfo {volume} {62}},\ \bibinfo
		{pages} {251} (\bibinfo {year} {1990})}\BibitemShut {NoStop}%
	\bibitem [{\citenamefont {Mel'nikov}(1991)}]{melnikov_fifty_years_kramers}%
	\BibitemOpen
	\bibfield  {author} {\bibinfo {author} {\bibfnamefont {V.}~\bibnamefont
			{Mel'nikov}},\ }\bibfield  {title} {\bibinfo {title} {The {K}ramers problem:
			Fifty years of development},\ }\href
	{https://doi.org/10.1016/0370-1573(91)90108-x} {\bibfield  {journal}
		{\bibinfo  {journal} {Phys. Rep.}\ }\textbf {\bibinfo {volume} {209}},\
		\bibinfo {pages} {1} (\bibinfo {year} {1991})}\BibitemShut {NoStop}%
	\bibitem [{\citenamefont {Marchetti}\ \emph {et~al.}(2013)\citenamefont
		{Marchetti}, \citenamefont {Joanny}, \citenamefont {Ramaswamy}, \citenamefont
		{Liverpool}, \citenamefont {Prost}, \citenamefont {Rao},\ and\ \citenamefont
		{Simha}}]{marchetti2013hydrodynamics}%
	\BibitemOpen
	\bibfield  {author} {\bibinfo {author} {\bibfnamefont {M.}~\bibnamefont
			{Marchetti}}, \bibinfo {author} {\bibfnamefont {J.}~\bibnamefont {Joanny}},
		\bibinfo {author} {\bibfnamefont {S.}~\bibnamefont {Ramaswamy}}, \bibinfo
		{author} {\bibfnamefont {T.}~\bibnamefont {Liverpool}}, \bibinfo {author}
		{\bibfnamefont {J.}~\bibnamefont {Prost}}, \bibinfo {author} {\bibfnamefont
			{M.}~\bibnamefont {Rao}},\ and\ \bibinfo {author} {\bibfnamefont {R.~A.}\
			\bibnamefont {Simha}},\ }\bibfield  {title} {\bibinfo {title} {Hydrodynamics
			of soft active matter},\ }\href {https://doi.org/10.1103/revmodphys.85.1143}
	{\bibfield  {journal} {\bibinfo  {journal} {Rev. Mod. Phys.}\ }\textbf
		{\bibinfo {volume} {85}},\ \bibinfo {pages} {1143} (\bibinfo {year}
		{2013})}\BibitemShut {NoStop}%
	\bibitem [{\citenamefont {Bechinger}\ \emph {et~al.}(2016)\citenamefont
		{Bechinger}, \citenamefont {Di~Leonardo}, \citenamefont {L\"owen},
		\citenamefont {Reichhardt}, \citenamefont {Volpe},\ and\ \citenamefont
		{Volpe}}]{Bechinger/etal:2016}%
	\BibitemOpen
	\bibfield  {author} {\bibinfo {author} {\bibfnamefont {C.}~\bibnamefont
			{Bechinger}}, \bibinfo {author} {\bibfnamefont {R.}~\bibnamefont
			{Di~Leonardo}}, \bibinfo {author} {\bibfnamefont {H.}~\bibnamefont
			{L\"owen}}, \bibinfo {author} {\bibfnamefont {C.}~\bibnamefont {Reichhardt}},
		\bibinfo {author} {\bibfnamefont {G.}~\bibnamefont {Volpe}},\ and\ \bibinfo
		{author} {\bibfnamefont {G.}~\bibnamefont {Volpe}},\ }\bibfield  {title}
	{\bibinfo {title} {Active particles in complex and crowded environments},\
	}\href {https://doi.org/10.1103/revmodphys.88.045006} {\bibfield  {journal}
		{\bibinfo  {journal} {Rev. Mod. Phys.}\ }\textbf {\bibinfo {volume} {88}},\
		\bibinfo {pages} {045006} (\bibinfo {year} {2016})}\BibitemShut {NoStop}%
	\bibitem [{\citenamefont {Levis}\ \emph {et~al.}(2017)\citenamefont {Levis},
		\citenamefont {Codina},\ and\ \citenamefont
		{Pagonabarraga}}]{levis2017active}%
	\BibitemOpen
	\bibfield  {author} {\bibinfo {author} {\bibfnamefont {D.}~\bibnamefont
			{Levis}}, \bibinfo {author} {\bibfnamefont {J.}~\bibnamefont {Codina}},\ and\
		\bibinfo {author} {\bibfnamefont {I.}~\bibnamefont {Pagonabarraga}},\
	}\bibfield  {title} {\bibinfo {title} {Active {B}rownian equation of state:
			metastability and phase coexistence},\ }\href
	{https://doi.org/10.1039/c7sm01504f} {\bibfield  {journal} {\bibinfo
			{journal} {Soft Matter}\ }\textbf {\bibinfo {volume} {13}},\ \bibinfo {pages}
		{8113} (\bibinfo {year} {2017})}\BibitemShut {NoStop}%
	\bibitem [{\citenamefont {te~Vrugt}\ \emph {et~al.}(2026)\citenamefont
		{te~Vrugt}, \citenamefont {Liebchen},\ and\ \citenamefont
		{Cates}}]{te2026colloquium}%
	\BibitemOpen
	\bibfield  {author} {\bibinfo {author} {\bibfnamefont {M.}~\bibnamefont
			{te~Vrugt}}, \bibinfo {author} {\bibfnamefont {B.}~\bibnamefont {Liebchen}},\
		and\ \bibinfo {author} {\bibfnamefont {M.~E.}\ \bibnamefont {Cates}},\
	}\bibfield  {title} {\bibinfo {title} {Colloquium: What do we mean by `active
			matter'?},\ }\href {https://doi.org/10.1103/wd4f-q7kv} {\bibfield  {journal}
		{\bibinfo  {journal} {Rev. Mod. Phys.}\ }\textbf {\bibinfo {volume} {98}},\
		\bibinfo {pages} {031001} (\bibinfo {year} {2026})}\BibitemShut {NoStop}%
	\bibitem [{\citenamefont {Elgeti}\ \emph {et~al.}(2015)\citenamefont {Elgeti},
		\citenamefont {Winkler},\ and\ \citenamefont {Gompper}}]{Elgeti/etal:2015}%
	\BibitemOpen
	\bibfield  {author} {\bibinfo {author} {\bibfnamefont {J.}~\bibnamefont
			{Elgeti}}, \bibinfo {author} {\bibfnamefont {R.~G.}\ \bibnamefont
			{Winkler}},\ and\ \bibinfo {author} {\bibfnamefont {G.}~\bibnamefont
			{Gompper}},\ }\bibfield  {title} {\bibinfo {title} {{{{{{{{{Physics}}}}}}} of
				microswimmers—single particle motion and collective behavior: a review}},\
	}\href {https://doi.org/10.1088/0034-4885/78/5/056601} {\bibfield  {journal}
		{\bibinfo  {journal} {Rep. Prog. Phys.}\ }\textbf {\bibinfo {volume} {78}},\
		\bibinfo {pages} {056601} (\bibinfo {year} {2015})}\BibitemShut {NoStop}%
	\bibitem [{\citenamefont {Burada}\ and\ \citenamefont
		{Lindner}(2012)}]{Burada/Lindner:2012}%
	\BibitemOpen
	\bibfield  {author} {\bibinfo {author} {\bibfnamefont {P.~S.}\ \bibnamefont
			{Burada}}\ and\ \bibinfo {author} {\bibfnamefont {B.}~\bibnamefont
			{Lindner}},\ }\bibfield  {title} {\bibinfo {title} {Escape rate of an active
			{B}rownian particle over a potential barrier},\ }\href
	{https://doi.org/10.1103/physreve.85.032102} {\bibfield  {journal} {\bibinfo
			{journal} {Phys. Rev. E}\ }\textbf {\bibinfo {volume} {85}},\ \bibinfo
		{pages} {032102} (\bibinfo {year} {2012})}\BibitemShut {NoStop}%
	\bibitem [{\citenamefont {Geiseler}\ \emph {et~al.}(2016)\citenamefont
		{Geiseler}, \citenamefont {H{\"a}nggi},\ and\ \citenamefont
		{Schmid}}]{geiseler2016kramers}%
	\BibitemOpen
	\bibfield  {author} {\bibinfo {author} {\bibfnamefont {A.}~\bibnamefont
			{Geiseler}}, \bibinfo {author} {\bibfnamefont {P.}~\bibnamefont
			{H{\"a}nggi}},\ and\ \bibinfo {author} {\bibfnamefont {G.}~\bibnamefont
			{Schmid}},\ }\bibfield  {title} {\bibinfo {title} {Kramers escape of a
			self-propelled particle},\ }\href
	{https://doi.org/10.1140/epjb/e2016-70359-0} {\bibfield  {journal} {\bibinfo
			{journal} {Eur. Phys. J. B}\ }\textbf {\bibinfo {volume} {89}},\ \bibinfo
		{pages} {175} (\bibinfo {year} {2016})}\BibitemShut {NoStop}%
	\bibitem [{\citenamefont {Koumakis}\ \emph {et~al.}(2016)\citenamefont
		{Koumakis}, \citenamefont {Gnoli}, \citenamefont {Maggi}, \citenamefont
		{Puglisi},\ and\ \citenamefont {Di~Leonardo}}]{Koumakis2016}%
	\BibitemOpen
	\bibfield  {author} {\bibinfo {author} {\bibfnamefont {N.}~\bibnamefont
			{Koumakis}}, \bibinfo {author} {\bibfnamefont {A.}~\bibnamefont {Gnoli}},
		\bibinfo {author} {\bibfnamefont {C.}~\bibnamefont {Maggi}}, \bibinfo
		{author} {\bibfnamefont {A.}~\bibnamefont {Puglisi}},\ and\ \bibinfo {author}
		{\bibfnamefont {R.}~\bibnamefont {Di~Leonardo}},\ }\bibfield  {title}
	{\bibinfo {title} {{{{{{Mechanism}}}} of self-propulsion in
				{{{{3D-printed}}}} active granular particles}},\ }\href
	{https://doi.org/10.1088/1367-2630/18/11/113046} {\bibfield  {journal}
		{\bibinfo  {journal} {New J. Phys.}\ }\textbf {\bibinfo {volume} {18}},\
		\bibinfo {pages} {113046} (\bibinfo {year} {2016})}\BibitemShut {NoStop}%
	\bibitem [{\citenamefont {Militaru}\ \emph {et~al.}(2021)\citenamefont
		{Militaru}, \citenamefont {Innerbichler}, \citenamefont {Frimmer},
		\citenamefont {Tebbenjohanns}, \citenamefont {Novotny},\ and\ \citenamefont
		{Dellago}}]{experiment_Militaru2021}%
	\BibitemOpen
	\bibfield  {author} {\bibinfo {author} {\bibfnamefont {A.}~\bibnamefont
			{Militaru}}, \bibinfo {author} {\bibfnamefont {M.}~\bibnamefont
			{Innerbichler}}, \bibinfo {author} {\bibfnamefont {M.}~\bibnamefont
			{Frimmer}}, \bibinfo {author} {\bibfnamefont {F.}~\bibnamefont
			{Tebbenjohanns}}, \bibinfo {author} {\bibfnamefont {L.}~\bibnamefont
			{Novotny}},\ and\ \bibinfo {author} {\bibfnamefont {C.}~\bibnamefont
			{Dellago}},\ }\bibfield  {title} {\bibinfo {title} {Escape dynamics of active
			particles in multistable potentials},\ }\href
	{https://doi.org/10.1038/s41467-021-22647-6} {\bibfield  {journal} {\bibinfo
			{journal} {Nat. Commun.}\ }\textbf {\bibinfo {volume} {12}},\ \bibinfo
		{pages} {2446} (\bibinfo {year} {2021})}\BibitemShut {NoStop}%
	\bibitem [{\citenamefont {Caprini}\ \emph {et~al.}(2019)\citenamefont
		{Caprini}, \citenamefont {Marini Bettolo~Marconi}, \citenamefont {Puglisi},\
		and\ \citenamefont {Vulpiani}}]{caprini2019active}%
	\BibitemOpen
	\bibfield  {author} {\bibinfo {author} {\bibfnamefont {L.}~\bibnamefont
			{Caprini}}, \bibinfo {author} {\bibfnamefont {U.}~\bibnamefont {Marini
				Bettolo~Marconi}}, \bibinfo {author} {\bibfnamefont {A.}~\bibnamefont
			{Puglisi}},\ and\ \bibinfo {author} {\bibfnamefont {A.}~\bibnamefont
			{Vulpiani}},\ }\bibfield  {title} {\bibinfo {title} {Active escape dynamics:
			The effect of persistence on barrier crossing},\ }\href
	{https://doi.org/10.1063/1.5080537} {\bibfield  {journal} {\bibinfo
			{journal} {J. Chem. Phys.}\ }\textbf {\bibinfo {volume} {150}},\ \bibinfo
		{pages} {024902} (\bibinfo {year} {2019})}\BibitemShut {NoStop}%
	\bibitem [{\citenamefont {Woillez}\ \emph {et~al.}(2019)\citenamefont
		{Woillez}, \citenamefont {Zhao}, \citenamefont {Kafri}, \citenamefont
		{Lecomte},\ and\ \citenamefont {Tailleur}}]{Woillez/etal:2019}%
	\BibitemOpen
	\bibfield  {author} {\bibinfo {author} {\bibfnamefont {E.}~\bibnamefont
			{Woillez}}, \bibinfo {author} {\bibfnamefont {Y.}~\bibnamefont {Zhao}},
		\bibinfo {author} {\bibfnamefont {Y.}~\bibnamefont {Kafri}}, \bibinfo
		{author} {\bibfnamefont {V.}~\bibnamefont {Lecomte}},\ and\ \bibinfo {author}
		{\bibfnamefont {J.}~\bibnamefont {Tailleur}},\ }\bibfield  {title} {\bibinfo
		{title} {Activated escape of a self-propelled particle from a metastable
			state},\ }\href {https://doi.org/10.1103/physrevlett.122.258001} {\bibfield
		{journal} {\bibinfo  {journal} {Phys. Rev. Lett.}\ }\textbf {\bibinfo
			{volume} {122}},\ \bibinfo {pages} {258001} (\bibinfo {year}
		{2019})}\BibitemShut {NoStop}%
	\bibitem [{\citenamefont {Baule}\ and\ \citenamefont
		{Sollich}(2023)}]{baule2023exponential}%
	\BibitemOpen
	\bibfield  {author} {\bibinfo {author} {\bibfnamefont {A.}~\bibnamefont
			{Baule}}\ and\ \bibinfo {author} {\bibfnamefont {P.}~\bibnamefont
			{Sollich}},\ }\bibfield  {title} {\bibinfo {title} {Exponential increase of
			transition rates in metastable systems driven by non-gaussian noise},\ }\href
	{https://doi.org/10.1038/s41598-023-30577-0} {\bibfield  {journal} {\bibinfo
			{journal} {Sci. Rep.}\ }\textbf {\bibinfo {volume} {13}},\ \bibinfo {pages}
		{3853} (\bibinfo {year} {2023})}\BibitemShut {NoStop}%
	\bibitem [{\citenamefont {Woillez}\ \emph {et~al.}(2020)\citenamefont
		{Woillez}, \citenamefont {Kafri},\ and\ \citenamefont
		{Lecomte}}]{woillez_nonlocal:2020}%
	\BibitemOpen
	\bibfield  {author} {\bibinfo {author} {\bibfnamefont {E.}~\bibnamefont
			{Woillez}}, \bibinfo {author} {\bibfnamefont {Y.}~\bibnamefont {Kafri}},\
		and\ \bibinfo {author} {\bibfnamefont {V.}~\bibnamefont {Lecomte}},\
	}\bibfield  {title} {\bibinfo {title} {Nonlocal stationary probability
			distributions and escape rates for an active {Ornstein--Uhlenbeck}
			particle},\ }\href {https://doi.org/10.1088/1742-5468/ab7e2e} {\bibfield
		{journal} {\bibinfo  {journal} {J. Stat. Mech.: Theory Exp.}\ }\textbf
		{\bibinfo {volume} {2020}},\ \bibinfo {pages} {063204} (\bibinfo {year} {2020})}\BibitemShut {NoStop}%
	\bibitem [{\citenamefont {Crisanti}\ and\ \citenamefont
		{Paoluzzi}(2026)}]{Crisanti/Paoluzzi_exactHam:2026}%
	\BibitemOpen
	\bibfield  {author} {\bibinfo {author} {\bibfnamefont {A.}~\bibnamefont
			{Crisanti}}\ and\ \bibinfo {author} {\bibfnamefont {M.}~\bibnamefont
			{Paoluzzi}},\ }\bibfield  {title} {\bibinfo {title} {Exact {H}amiltonian
			dynamics of rare events in active matter},\ }\href
	{https://doi.org/10.48550/arXiv.2609.07496} {\bibfield  {journal} {\bibinfo
			{journal} {arXiv:2609.07496}\ } (\bibinfo {year} {2026})}\BibitemShut
	{NoStop}%
	\bibitem [{\citenamefont {Olsen}\ \emph {et~al.}(2020)\citenamefont {Olsen},
		\citenamefont {Angheluta},\ and\ \citenamefont
		{Flekk\o{}y}}]{Olsen_disk2020}%
	\BibitemOpen
	\bibfield  {author} {\bibinfo {author} {\bibfnamefont {K.~S.}\ \bibnamefont
			{Olsen}}, \bibinfo {author} {\bibfnamefont {L.}~\bibnamefont {Angheluta}},\
		and\ \bibinfo {author} {\bibfnamefont {E.~G.}\ \bibnamefont {Flekk\o{}y}},\
	}\bibfield  {title} {\bibinfo {title} {Escape problem for active particles
			confined to a disk},\ }\href
	{https://doi.org/10.1103/physrevresearch.2.043314} {\bibfield  {journal}
		{\bibinfo  {journal} {Phys. Rev. Res.}\ }\textbf {\bibinfo {volume} {2}},\
		\bibinfo {pages} {043314} (\bibinfo {year} {2020})}\BibitemShut {NoStop}%
	\bibitem [{\citenamefont {Wexler}\ \emph {et~al.}(2020)\citenamefont {Wexler},
		\citenamefont {Gov}, \citenamefont {Rasmussen},\ and\ \citenamefont
		{Bel}}]{harmonic_Wexler2020}%
	\BibitemOpen
	\bibfield  {author} {\bibinfo {author} {\bibfnamefont {D.}~\bibnamefont
			{Wexler}}, \bibinfo {author} {\bibfnamefont {N.}~\bibnamefont {Gov}},
		\bibinfo {author} {\bibfnamefont {K.~O.}\ \bibnamefont {Rasmussen}},\ and\
		\bibinfo {author} {\bibfnamefont {G.}~\bibnamefont {Bel}},\ }\bibfield
	{title} {\bibinfo {title} {Dynamics and escape of active particles in a
			harmonic trap},\ }\href {https://doi.org/10.1103/physrevresearch.2.013003}
	{\bibfield  {journal} {\bibinfo  {journal} {Phys. Rev. Res.}\ }\textbf
		{\bibinfo {volume} {2}},\ \bibinfo {pages} {013003} (\bibinfo {year}
		{2020})}\BibitemShut {NoStop}%
	\bibitem [{\citenamefont {Zanovello}\ \emph {et~al.}(2021)\citenamefont
		{Zanovello}, \citenamefont {Caraglio}, \citenamefont {Franosch},\ and\
		\citenamefont {Faccioli}}]{Zanovello/etal:2021}%
	\BibitemOpen
	\bibfield  {author} {\bibinfo {author} {\bibfnamefont {L.}~\bibnamefont
			{Zanovello}}, \bibinfo {author} {\bibfnamefont {M.}~\bibnamefont {Caraglio}},
		\bibinfo {author} {\bibfnamefont {T.}~\bibnamefont {Franosch}},\ and\
		\bibinfo {author} {\bibfnamefont {P.}~\bibnamefont {Faccioli}},\ }\bibfield
	{title} {\bibinfo {title} {Target search of active agents crossing high
			energy barriers},\ }\href {https://doi.org/10.1103/PhysRevLett.126.018001}
	{\bibfield  {journal} {\bibinfo  {journal} {Phys. Rev. Lett.}\ }\textbf
		{\bibinfo {volume} {126}},\ \bibinfo {pages} {018001} (\bibinfo {year}
		{2021})}\BibitemShut {NoStop}%
	\bibitem [{\citenamefont {Aranson}\ and\ \citenamefont
		{Pikovsky}(2022)}]{collective_Aranson2022}%
	\BibitemOpen
	\bibfield  {author} {\bibinfo {author} {\bibfnamefont {I.~S.}\ \bibnamefont
			{Aranson}}\ and\ \bibinfo {author} {\bibfnamefont {A.}~\bibnamefont
			{Pikovsky}},\ }\bibfield  {title} {\bibinfo {title} {Confinement and
			collective escape of active particles},\ }\href
	{https://doi.org/10.1103/physrevlett.128.108001} {\bibfield  {journal}
		{\bibinfo  {journal} {Phys. Rev. Lett.}\ }\textbf {\bibinfo {volume} {128}},\
		\bibinfo {pages} {108001} (\bibinfo {year} {2022})}\BibitemShut {NoStop}%
	\bibitem [{\citenamefont {Tasinkevych}\ \emph {et~al.}(2026)\citenamefont
		{Tasinkevych}, \citenamefont {Le},\ and\ \citenamefont
		{Ryabov}}]{Tasinkevych/etal:2026}%
	\BibitemOpen
	\bibfield  {author} {\bibinfo {author} {\bibfnamefont {M.}~\bibnamefont
			{Tasinkevych}}, \bibinfo {author} {\bibfnamefont {X.~M.}\ \bibnamefont
			{Le}},\ and\ \bibinfo {author} {\bibfnamefont {A.}~\bibnamefont {Ryabov}},\
	}\href {https://arxiv.org/abs/2609.27425} {\bibinfo {title} {Uphill and
			downhill first passage of an active brownian particle: Asymmetry and exact
			path reweighting}} (\bibinfo {year} {2026}),\ \Eprint
	{https://arxiv.org/abs/arxiv:2609.27425} {arxiv:2609.27425} \BibitemShut
	{NoStop}%
	\bibitem [{\citenamefont {Miao}\ \emph {et~al.}(2025)\citenamefont {Miao},
		\citenamefont {Liu}, \citenamefont {She}, \citenamefont {Li}, \citenamefont
		{Bao},\ and\ \citenamefont {Li}}]{Lin_ABP2025}%
	\BibitemOpen
	\bibfield  {author} {\bibinfo {author} {\bibfnamefont {L.}~\bibnamefont
			{Miao}}, \bibinfo {author} {\bibfnamefont {J.}~\bibnamefont {Liu}}, \bibinfo
		{author} {\bibfnamefont {H.-Z.}\ \bibnamefont {She}}, \bibinfo {author}
		{\bibfnamefont {P.-C.}\ \bibnamefont {Li}}, \bibinfo {author} {\bibfnamefont
			{J.-D.}\ \bibnamefont {Bao}},\ and\ \bibinfo {author} {\bibfnamefont {M.-G.}\
			\bibnamefont {Li}},\ }\bibfield  {title} {\bibinfo {title} {Escape dynamics
			of active {B}rownian particles with multimodal diffusion},\ }\href
	{https://doi.org/10.1063/5.0297646} {\bibfield  {journal} {\bibinfo
			{journal} {J. Chem. Phys.}\ }\textbf {\bibinfo {volume} {163}},\ \bibinfo
		{pages} {174907} (\bibinfo {year} {2025})}\BibitemShut {NoStop}%
	\bibitem [{\citenamefont {Basu}\ \emph {et~al.}(2025)\citenamefont {Basu},
		\citenamefont {Majumdar},\ and\ \citenamefont {Rosso}}]{RTP_Basu2026}%
	\BibitemOpen
	\bibfield  {author} {\bibinfo {author} {\bibfnamefont {U.}~\bibnamefont
			{Basu}}, \bibinfo {author} {\bibfnamefont {S.~N.}\ \bibnamefont {Majumdar}},\
		and\ \bibinfo {author} {\bibfnamefont {A.}~\bibnamefont {Rosso}},\ }\bibfield
	{title} {\bibinfo {title} {Ergodicity breaking in active run-and-tumble
			particles in a double-well potential},\ }\href
	{https://doi.org/10.1103/jysw-w2mb} {\bibfield  {journal} {\bibinfo
			{journal} {APS Open Sci.}\ }\textbf {\bibinfo {volume} {1}},\ \bibinfo
		{pages} {000021} (\bibinfo {year} {2025})}\BibitemShut {NoStop}%
	\bibitem [{\citenamefont {Nayak}\ \emph {et~al.}(2026)\citenamefont {Nayak},
		\citenamefont {Bag}, \citenamefont {Bhattacharyya}, \citenamefont {Deb},
		\citenamefont {Mandal},\ and\ \citenamefont {Ghosh}}]{Huyak/etal:2026}%
	\BibitemOpen
	\bibfield  {author} {\bibinfo {author} {\bibfnamefont {S.}~\bibnamefont
			{Nayak}}, \bibinfo {author} {\bibfnamefont {P.}~\bibnamefont {Bag}}, \bibinfo
		{author} {\bibfnamefont {P.}~\bibnamefont {Bhattacharyya}}, \bibinfo {author}
		{\bibfnamefont {S.}~\bibnamefont {Deb}}, \bibinfo {author} {\bibfnamefont
			{D.}~\bibnamefont {Mandal}},\ and\ \bibinfo {author} {\bibfnamefont {P.~K.}\
			\bibnamefont {Ghosh}},\ }\bibfield  {title} {\bibinfo {title} {Escape
			dynamics of elliptical {B}rownian particles from cavities: Numerical
			simulations},\ }\href {https://doi.org/10.1021/acs.jpcb.6c01287} {\bibfield
		{journal} {\bibinfo  {journal} {J. Phys. Chem. B}\ }\textbf {\bibinfo
			{volume} {130}},\ \bibinfo {pages} {5420} (\bibinfo {year}
		{2026})}\BibitemShut {NoStop}%
	\bibitem [{\citenamefont {Wei}\ \emph {et~al.}(2026)\citenamefont {Wei},
		\citenamefont {Hu}, \citenamefont {Chen}, \citenamefont {Dai}, \citenamefont
		{Jiao}, \citenamefont {Meng},\ and\ \citenamefont {Yan}}]{Wei2026}%
	\BibitemOpen
	\bibfield  {author} {\bibinfo {author} {\bibfnamefont {W.}~\bibnamefont
			{Wei}}, \bibinfo {author} {\bibfnamefont {S.}~\bibnamefont {Hu}}, \bibinfo
		{author} {\bibfnamefont {W.}~\bibnamefont {Chen}}, \bibinfo {author}
		{\bibfnamefont {X.}~\bibnamefont {Dai}}, \bibinfo {author} {\bibfnamefont
			{Z.}~\bibnamefont {Jiao}}, \bibinfo {author} {\bibfnamefont {F.}~\bibnamefont
			{Meng}},\ and\ \bibinfo {author} {\bibfnamefont {L.-T.}\ \bibnamefont
			{Yan}},\ }\bibfield  {title} {\bibinfo {title} {Hydrodynamically controlled
			active escape dynamics},\ }\href {https://doi.org/10.1103/pzsy-5k17}
	{\bibfield  {journal} {\bibinfo  {journal} {Phys. Rev. Lett.}\ }\textbf
		{\bibinfo {volume} {136}},\ \bibinfo {pages} {128301} (\bibinfo {year}
		{2026})}\BibitemShut {NoStop}%
	\bibitem [{\citenamefont {Gu\'eneau}\ \emph {et~al.}(2025)\citenamefont
		{Gu\'eneau}, \citenamefont {Majumdar},\ and\ \citenamefont
		{Schehr}}]{Gueneau/Majumdar:2025}%
	\BibitemOpen
	\bibfield  {author} {\bibinfo {author} {\bibfnamefont {M.}~\bibnamefont
			{Gu\'eneau}}, \bibinfo {author} {\bibfnamefont {S.~N.}\ \bibnamefont
			{Majumdar}},\ and\ \bibinfo {author} {\bibfnamefont {G.}~\bibnamefont
			{Schehr}},\ }\bibfield  {title} {\bibinfo {title} {Run-and-tumble particle in
			one-dimensional potentials: Mean first-passage time and applications},\
	}\href {https://doi.org/10.1103/PhysRevE.111.014144} {\bibfield  {journal}
		{\bibinfo  {journal} {Phys. Rev. E}\ }\textbf {\bibinfo {volume} {111}},\
		\bibinfo {pages} {014144} (\bibinfo {year} {2025})}\BibitemShut {NoStop}%
	\bibitem [{\citenamefont {Sharma}\ \emph {et~al.}(2017)\citenamefont {Sharma},
		\citenamefont {Wittmann},\ and\ \citenamefont {Brader}}]{Sharma/etal:2017}%
	\BibitemOpen
	\bibfield  {author} {\bibinfo {author} {\bibfnamefont {A.}~\bibnamefont
			{Sharma}}, \bibinfo {author} {\bibfnamefont {R.}~\bibnamefont {Wittmann}},\
		and\ \bibinfo {author} {\bibfnamefont {J.~M.}\ \bibnamefont {Brader}},\
	}\bibfield  {title} {\bibinfo {title} {Escape rate of active particles in the
			effective equilibrium approach},\ }\href
	{https://doi.org/10.1103/physreve.95.012115} {\bibfield  {journal} {\bibinfo
			{journal} {Phys. Rev. E}\ }\textbf {\bibinfo {volume} {95}},\ \bibinfo
		{pages} {012115} (\bibinfo {year} {2017})}\BibitemShut {NoStop}%
	\bibitem [{\citenamefont {Chaki}\ and\ \citenamefont
		{Chakrabarti}(2020)}]{Chaki2020}%
	\BibitemOpen
	\bibfield  {author} {\bibinfo {author} {\bibfnamefont {S.}~\bibnamefont
			{Chaki}}\ and\ \bibinfo {author} {\bibfnamefont {R.}~\bibnamefont
			{Chakrabarti}},\ }\bibfield  {title} {\bibinfo {title} {Escape of a passive
			particle from an activity-induced energy landscape: emergence of slow and
			fast effective diffusion},\ }\href {https://doi.org/10.1039/d0sm00711k}
	{\bibfield  {journal} {\bibinfo  {journal} {Soft Matter}\ }\textbf {\bibinfo
			{volume} {16}},\ \bibinfo {pages} {7103} (\bibinfo {year}
		{2020})}\BibitemShut {NoStop}%
	\bibitem [{\citenamefont {Caprini}\ \emph
		{et~al.}(2021{\natexlab{a}})\citenamefont {Caprini}, \citenamefont
		{Cecconi},\ and\ \citenamefont {Marini
			Bettolo~Marconi}}]{caprini2021correlated}%
	\BibitemOpen
	\bibfield  {author} {\bibinfo {author} {\bibfnamefont {L.}~\bibnamefont
			{Caprini}}, \bibinfo {author} {\bibfnamefont {F.}~\bibnamefont {Cecconi}},\
		and\ \bibinfo {author} {\bibfnamefont {U.}~\bibnamefont {Marini
				Bettolo~Marconi}},\ }\bibfield  {title} {\bibinfo {title} {Correlated escape
			of active particles across a potential barrier},\ }\href
	{https://doi.org/10.1063/5.0074072} {\bibfield  {journal} {\bibinfo
			{journal} {J. Chem. Phys.}\ }\textbf {\bibinfo {volume} {155}},\ \bibinfo
		{pages} {234902} (\bibinfo {year} {2021}{\natexlab{a}})}\BibitemShut
	{NoStop}%
	\bibitem [{\citenamefont {Castro-Villarreal}\ and\ \citenamefont
		{Sevilla}(2018)}]{Castro/Sevilla:2018}%
	\BibitemOpen
	\bibfield  {author} {\bibinfo {author} {\bibfnamefont {P.}~\bibnamefont
			{Castro-Villarreal}}\ and\ \bibinfo {author} {\bibfnamefont {F.~J.}\
			\bibnamefont {Sevilla}},\ }\bibfield  {title} {\bibinfo {title} {Active
			motion on curved surfaces},\ }\href
	{https://doi.org/10.1103/physreve.97.052605} {\bibfield  {journal} {\bibinfo
			{journal} {Phys. Rev. E}\ }\textbf {\bibinfo {volume} {97}},\ \bibinfo
		{pages} {052605} (\bibinfo {year} {2018})}\BibitemShut {NoStop}%
	\bibitem [{\citenamefont {Castro-Villarreal}\ \emph {et~al.}(2023)\citenamefont
		{Castro-Villarreal}, \citenamefont {Solano-Cabrera},\ and\ \citenamefont
		{Castañeda-Priego}}]{castro2023}%
	\BibitemOpen
	\bibfield  {author} {\bibinfo {author} {\bibfnamefont {P.}~\bibnamefont
			{Castro-Villarreal}}, \bibinfo {author} {\bibfnamefont {C.~O.}\ \bibnamefont
			{Solano-Cabrera}},\ and\ \bibinfo {author} {\bibfnamefont {R.}~\bibnamefont
			{Castañeda-Priego}},\ }\bibfield  {title} {\bibinfo {title} {Covariant
			description of the colloidal dynamics on curved manifolds},\ }\href
	{https://doi.org/10.3389/fphy.2023.1204751} {\bibfield  {journal} {\bibinfo
			{journal} {Front. Phys.}\ }\textbf {\bibinfo {volume} {11}},\ \bibinfo
		{pages} {1204751} (\bibinfo {year} {2023})}\BibitemShut {NoStop}%
	\bibitem [{\citenamefont {Iyer}\ \emph {et~al.}(2023)\citenamefont {Iyer},
		\citenamefont {Winkler}, \citenamefont {Fedosov},\ and\ \citenamefont
		{Gompper}}]{Iyer/etal:2023}%
	\BibitemOpen
	\bibfield  {author} {\bibinfo {author} {\bibfnamefont {P.}~\bibnamefont
			{Iyer}}, \bibinfo {author} {\bibfnamefont {R.~G.}\ \bibnamefont {Winkler}},
		\bibinfo {author} {\bibfnamefont {D.~A.}\ \bibnamefont {Fedosov}},\ and\
		\bibinfo {author} {\bibfnamefont {G.}~\bibnamefont {Gompper}},\ }\bibfield
	{title} {\bibinfo {title} {Dynamics and phase separation of active {B}rownian
			particles on curved surfaces and in porous media},\ }\href
	{https://doi.org/10.1103/physrevresearch.5.033054} {\bibfield  {journal}
		{\bibinfo  {journal} {Phys. Rev. Res.}\ }\textbf {\bibinfo {volume} {5}},\
		\bibinfo {pages} {033054} (\bibinfo {year} {2023})}\BibitemShut {NoStop}%
	\bibitem [{\citenamefont {Mackay}\ \emph {et~al.}(2026)\citenamefont {Mackay},
		\citenamefont {Janzen}, \citenamefont {Matoz-Fernandez},\ and\ \citenamefont
		{Sknepnek}}]{Mackay/etal:2026}%
	\BibitemOpen
	\bibfield  {author} {\bibinfo {author} {\bibfnamefont {E.~D.}\ \bibnamefont
			{Mackay}}, \bibinfo {author} {\bibfnamefont {G.}~\bibnamefont {Janzen}},
		\bibinfo {author} {\bibfnamefont {D.~A.}\ \bibnamefont {Matoz-Fernandez}},\
		and\ \bibinfo {author} {\bibfnamefont {R.}~\bibnamefont {Sknepnek}},\
	}\bibfield  {title} {\bibinfo {title} {Emergent dynamics of active systems on
			curved environments},\ }\href {https://doi.org/10.1103/t9rb-blzc} {\bibfield
		{journal} {\bibinfo  {journal} {Phys. Rev. Res.}\ }\textbf {\bibinfo {volume}
			{8}},\ \bibinfo {pages} {013172} (\bibinfo {year} {2026})}\BibitemShut
	{NoStop}%
	\bibitem [{\citenamefont {Caprini}\ and\ \citenamefont
		{Marconi}(2018)}]{Caprini/Marconi:2018}%
	\BibitemOpen
	\bibfield  {author} {\bibinfo {author} {\bibfnamefont {L.}~\bibnamefont
			{Caprini}}\ and\ \bibinfo {author} {\bibfnamefont {U.~M.~B.}\ \bibnamefont
			{Marconi}},\ }\bibfield  {title} {\bibinfo {title} {Active particles under
			confinement and effective force generation among surfaces},\ }\href
	{https://doi.org/10.1039/c8sm01840e} {\bibfield  {journal} {\bibinfo
			{journal} {Soft Matter}\ }\textbf {\bibinfo {volume} {14}},\ \bibinfo {pages}
		{9044} (\bibinfo {year} {2018})}\BibitemShut {NoStop}%
	\bibitem [{\citenamefont {Apaza}\ and\ \citenamefont
		{Sandoval}(2018)}]{Sandoval2018}%
	\BibitemOpen
	\bibfield  {author} {\bibinfo {author} {\bibfnamefont {L.}~\bibnamefont
			{Apaza}}\ and\ \bibinfo {author} {\bibfnamefont {M.}~\bibnamefont
			{Sandoval}},\ }\bibfield  {title} {\bibinfo {title} {Active matter on
			{R}iemannian manifolds},\ }\href {https://doi.org/10.1039/c8sm01034j}
	{\bibfield  {journal} {\bibinfo  {journal} {Soft Matter}\ }\textbf {\bibinfo
			{volume} {14}},\ \bibinfo {pages} {9928} (\bibinfo {year}
		{2018})}\BibitemShut {NoStop}%
	\bibitem [{\citenamefont {Webb}\ \emph {et~al.}(2026)\citenamefont {Webb},
		\citenamefont {Ansell},\ and\ \citenamefont
		{Sussman}}]{Webb/etal_MIPS_curved:2026}%
	\BibitemOpen
	\bibfield  {author} {\bibinfo {author} {\bibfnamefont {T.~H.}\ \bibnamefont
			{Webb}}, \bibinfo {author} {\bibfnamefont {H.~S.}\ \bibnamefont {Ansell}},\
		and\ \bibinfo {author} {\bibfnamefont {D.~M.}\ \bibnamefont {Sussman}},\
	}\bibfield  {title} {\bibinfo {title} {Geometric control of motility-induced
			phase separation},\ }\href {https://doi.org/10.1039/d6sm00213g} {\bibfield
		{journal} {\bibinfo  {journal} {Soft Matter}\ }\textbf {\bibinfo {volume}
			{22}},\ \bibinfo {pages} {4082} (\bibinfo {year} {2026})}\BibitemShut
	{NoStop}%
	\bibitem [{\citenamefont {Naji}\ and\ \citenamefont
		{Brown}(2007)}]{Naji/Brown2007}%
	\BibitemOpen
	\bibfield  {author} {\bibinfo {author} {\bibfnamefont {A.}~\bibnamefont
			{Naji}}\ and\ \bibinfo {author} {\bibfnamefont {F.~L.~H.}\ \bibnamefont
			{Brown}},\ }\bibfield  {title} {\bibinfo {title} {Diffusion on ruffled
			membrane surfaces},\ }\href {https://doi.org/10.1063/1.2739526} {\bibfield
		{journal} {\bibinfo  {journal} {J. Chem. Phys.}\ }\textbf {\bibinfo {volume}
			{126}},\ \bibinfo {pages} {235103} (\bibinfo {year} {2007})}\BibitemShut
	{NoStop}%
	\bibitem [{\citenamefont {Ohta}\ and\ \citenamefont
		{Komura}(2020)}]{Ohta/Komura2020}%
	\BibitemOpen
	\bibfield  {author} {\bibinfo {author} {\bibfnamefont {T.}~\bibnamefont
			{Ohta}}\ and\ \bibinfo {author} {\bibfnamefont {S.}~\bibnamefont {Komura}},\
	}\bibfield  {title} {\bibinfo {title} {Lateral diffusion on a frozen random
			surface},\ }\href {https://doi.org/10.1209/0295-5075/132/50007} {\bibfield
		{journal} {\bibinfo  {journal} {EPL}\ }\textbf {\bibinfo {volume} {132}},\
		\bibinfo {pages} {50007} (\bibinfo {year} {2020})}\BibitemShut {NoStop}%
	\bibitem [{\citenamefont {Fily}\ \emph {et~al.}(2016)\citenamefont {Fily},
		\citenamefont {Baskaran},\ and\ \citenamefont {Hagan}}]{Fily/Baskaran:2016}%
	\BibitemOpen
	\bibfield  {author} {\bibinfo {author} {\bibfnamefont {Y.}~\bibnamefont
			{Fily}}, \bibinfo {author} {\bibfnamefont {A.}~\bibnamefont {Baskaran}},\
		and\ \bibinfo {author} {\bibfnamefont {M.~F.}\ \bibnamefont {Hagan}},\
	}\bibfield  {title} {\bibinfo {title} {Active particles on curved surfaces},\
	}\href {https://doi.org/10.48550/arXiv.1601.00324} {\bibfield  {journal}
		{\bibinfo  {journal} {arXiv:1601.00324}\ } (\bibinfo {year}
		{2016})}\BibitemShut {NoStop}%
	\bibitem [{\citenamefont {Li}\ \emph {et~al.}(2024)\citenamefont {Li},
		\citenamefont {Liu},\ and\ \citenamefont {Wang}}]{Li_polar_active:2024}%
	\BibitemOpen
	\bibfield  {author} {\bibinfo {author} {\bibfnamefont {J.}~\bibnamefont
			{Li}}, \bibinfo {author} {\bibfnamefont {C.}~\bibnamefont {Liu}},\ and\
		\bibinfo {author} {\bibfnamefont {Q.}~\bibnamefont {Wang}},\ }\bibfield
	{title} {\bibinfo {title} {Emergent dynamics: Collective motions of polar
			active particles on surfaces},\ }\href {https://doi.org/10.1063/5.0204339}
	{\bibfield  {journal} {\bibinfo  {journal} {Phys. Fluids}\ }\textbf {\bibinfo
			{volume} {36}},\ \bibinfo {pages} {061907} (\bibinfo {year}
		{2024})}\BibitemShut {NoStop}%
	\bibitem [{\citenamefont {Sknepnek}\ and\ \citenamefont
		{Henkes}(2015)}]{Sknepnek/Henkes:2015}%
	\BibitemOpen
	\bibfield  {author} {\bibinfo {author} {\bibfnamefont {R.}~\bibnamefont
			{Sknepnek}}\ and\ \bibinfo {author} {\bibfnamefont {S.}~\bibnamefont
			{Henkes}},\ }\bibfield  {title} {\bibinfo {title} {Active swarms on a
			sphere},\ }\href {https://doi.org/10.1103/PhysRevE.91.022306} {\bibfield
		{journal} {\bibinfo  {journal} {Phys. Rev. E}\ }\textbf {\bibinfo {volume}
			{91}},\ \bibinfo {pages} {022306} (\bibinfo {year} {2015})}\BibitemShut
	{NoStop}%
	\bibitem [{\citenamefont {Caprini}\ \emph
		{et~al.}(2021{\natexlab{b}})\citenamefont {Caprini}, \citenamefont {Maggi},\
		and\ \citenamefont {Marini Bettolo~Marconi}}]{Caprini/etal:2021}%
	\BibitemOpen
	\bibfield  {author} {\bibinfo {author} {\bibfnamefont {L.}~\bibnamefont
			{Caprini}}, \bibinfo {author} {\bibfnamefont {C.}~\bibnamefont {Maggi}},\
		and\ \bibinfo {author} {\bibfnamefont {U.}~\bibnamefont {Marini
				Bettolo~Marconi}},\ }\bibfield  {title} {\bibinfo {title} {Collective effects
			in confined active {B}rownian particles},\ }\href
	{https://doi.org/10.1063/5.0051315} {\bibfield  {journal} {\bibinfo
			{journal} {J. Chem. Phys.}\ }\textbf {\bibinfo {volume} {154}},\ \bibinfo
		{pages} {244901} (\bibinfo {year} {2021}{\natexlab{b}})}\BibitemShut
	{NoStop}%
	\bibitem [{\citenamefont {Nakahara}(2018)}]{Nakahara}%
	\BibitemOpen
	\bibfield  {author} {\bibinfo {author} {\bibfnamefont {M.}~\bibnamefont
			{Nakahara}},\ }\href {https://doi.org/10.1201/9781315275826} {\emph {\bibinfo
			{title} {Geometry, Topology and Physics}}}\ (\bibinfo  {publisher} {Taylor \&
		Francis},\ \bibinfo {year} {2018})\BibitemShut {NoStop}%
	\bibitem [{\citenamefont {Gro{\ss}mann}\ \emph {et~al.}(2015)\citenamefont
		{Gro{\ss}mann}, \citenamefont {Peruani},\ and\ \citenamefont
		{B{\"a}r}}]{Grossmann2015}%
	\BibitemOpen
	\bibfield  {author} {\bibinfo {author} {\bibfnamefont {R.}~\bibnamefont
			{Gro{\ss}mann}}, \bibinfo {author} {\bibfnamefont {F.}~\bibnamefont
			{Peruani}},\ and\ \bibinfo {author} {\bibfnamefont {M.}~\bibnamefont
			{B{\"a}r}},\ }\bibfield  {title} {\bibinfo {title} {A geometric approach to
			self-propelled motion in isotropic {\&} anisotropic environments},\ }\href
	{https://doi.org/10.1140/epjst/e2015-02465-0} {\bibfield  {journal} {\bibinfo
			{journal} {Eur. Phys. J. Spec. Top.}\ }\textbf {\bibinfo {volume} {224}},\
		\bibinfo {pages} {1377} (\bibinfo {year} {2015})}\BibitemShut {NoStop}%
	\bibitem [{\citenamefont {N\'emeth}\ and\ \citenamefont
		{Adhikari}(2025)}]{Nemeth/Adhikari:2025}%
	\BibitemOpen
	\bibfield  {author} {\bibinfo {author} {\bibfnamefont {B.}~\bibnamefont
			{N\'emeth}}\ and\ \bibinfo {author} {\bibfnamefont {R.}~\bibnamefont
			{Adhikari}},\ }\bibfield  {title} {\bibinfo {title} {Intrinsic {L}angevin
			dynamics of rigid inclusions on curved surfaces},\ }\href
	{https://doi.org/10.1103/PhysRevE.111.045418} {\bibfield  {journal} {\bibinfo
			{journal} {Phys. Rev. E}\ }\textbf {\bibinfo {volume} {111}},\ \bibinfo
		{pages} {045418} (\bibinfo {year} {2025})}\BibitemShut {NoStop}%
	\bibitem [{\citenamefont {Risken}(1984)}]{Risken1988}%
	\BibitemOpen
	\bibfield  {author} {\bibinfo {author} {\bibfnamefont {H.}~\bibnamefont
			{Risken}},\ }\href {https://doi.org/10.1007/978-3-642-96807-5} {\emph
		{\bibinfo {title} {The {{{{Fokker-Planck}}}} Equation}}}\ (\bibinfo
	{publisher} {Springer Berlin, Heidelberg},\ \bibinfo {year} {1984})\BibitemShut {NoStop}%
	\bibitem [{\citenamefont {van Kampen}(2007)}]{vanKampen}%
	\BibitemOpen
	\bibfield  {author} {\bibinfo {author} {\bibfnamefont {N.~G.}\ \bibnamefont
			{van Kampen}},\ }\href {https://doi.org/10.1016/B978-0-444-52965-7.X5000-4}
	{\emph {\bibinfo {title} {Stochastic Processes in Physics and Chemistry}}}\
	(\bibinfo  {publisher} {North-Holland Personal Library, Elsevier Science},\
	\bibinfo {address} {New York},\ \bibinfo {year} {2007})\BibitemShut {NoStop}%
	\bibitem [{\citenamefont {Ross}(1984)}]{Ross1996}%
	\BibitemOpen
	\bibfield  {author} {\bibinfo {author} {\bibfnamefont {S.~M.}\ \bibnamefont
			{Ross}},\ }\href@noop {} {\emph {\bibinfo {title} {Stochastic Processes}}},\
	\bibinfo {edition} {2nd}\ ed.,\ Vol.~\bibinfo {volume} {79}\ (\bibinfo
	{publisher} {Wiley},\ \bibinfo {address} {New York},\ \bibinfo {year}
	{1984})\BibitemShut {NoStop}%
	\bibitem [{\citenamefont {Gardiner}(1983)}]{gardiner2009stochastic}%
	\BibitemOpen
	\bibfield  {author} {\bibinfo {author} {\bibfnamefont {C.}~\bibnamefont
			{Gardiner}},\ }\href@noop {} {\emph {\bibinfo {title} {Stochastic
				methods}}},\ Vol.~\bibinfo {volume} {4}\ (\bibinfo  {publisher} {Springer
		Berlin Heidelberg},\ \bibinfo {year} {1983})\BibitemShut {NoStop}%
	\bibitem [{\citenamefont {Arnoldi}(1951)}]{arnoldi1951principle}%
	\BibitemOpen
	\bibfield  {author} {\bibinfo {author} {\bibfnamefont {W.~E.}\ \bibnamefont
			{Arnoldi}},\ }\bibfield  {title} {\bibinfo {title} {The principle of
			minimized iterations in the solution of the matrix eigenvalue problem},\
	}\href {https://doi.org/10.1090/qam/42792} {\bibfield  {journal} {\bibinfo
			{journal} {Q. Appl. Math}\ }\textbf {\bibinfo {volume} {9}},\ \bibinfo
		{pages} {17} (\bibinfo {year} {1951})}\BibitemShut {NoStop}%
	\bibitem [{\citenamefont {Palagi}\ and\ \citenamefont
		{Fischer}(2018)}]{palagi2018bioinspired}%
	\BibitemOpen
	\bibfield  {author} {\bibinfo {author} {\bibfnamefont {S.}~\bibnamefont
			{Palagi}}\ and\ \bibinfo {author} {\bibfnamefont {P.}~\bibnamefont
			{Fischer}},\ }\bibfield  {title} {\bibinfo {title} {Bioinspired
			microrobots},\ }\href {https://doi.org/10.1038/s41578-018-0016-9} {\bibfield
		{journal} {\bibinfo  {journal} {Nat. Rev. Mater.}\ }\textbf {\bibinfo
			{volume} {3}},\ \bibinfo {pages} {113} (\bibinfo {year} {2018})}\BibitemShut
	{NoStop}%
	\bibitem [{\citenamefont {Cichos}\ \emph {et~al.}(2020)\citenamefont {Cichos},
		\citenamefont {Gustavsson}, \citenamefont {Mehlig},\ and\ \citenamefont
		{Volpe}}]{cichos2020machine}%
	\BibitemOpen
	\bibfield  {author} {\bibinfo {author} {\bibfnamefont {F.}~\bibnamefont
			{Cichos}}, \bibinfo {author} {\bibfnamefont {K.}~\bibnamefont {Gustavsson}},
		\bibinfo {author} {\bibfnamefont {B.}~\bibnamefont {Mehlig}},\ and\ \bibinfo
		{author} {\bibfnamefont {G.}~\bibnamefont {Volpe}},\ }\bibfield  {title}
	{\bibinfo {title} {Machine learning for active matter},\ }\href
	{https://doi.org/10.1038/s42256-020-0146-9} {\bibfield  {journal} {\bibinfo
			{journal} {Nat. Mach. Intell.}\ }\textbf {\bibinfo {volume} {2}},\ \bibinfo
		{pages} {94} (\bibinfo {year} {2020})}\BibitemShut {NoStop}%
	\bibitem [{\citenamefont {L{\"o}wen}\ and\ \citenamefont
		{Liebchen}(2026)}]{lowen2026towards}%
	\BibitemOpen
	\bibfield  {author} {\bibinfo {author} {\bibfnamefont {H.}~\bibnamefont
			{L{\"o}wen}}\ and\ \bibinfo {author} {\bibfnamefont {B.}~\bibnamefont
			{Liebchen}},\ }\bibfield  {title} {\bibinfo {title} {Towards intelligent
			active particles},\ }in\ \href
	{https://doi.org/10.1007/978-3-032-04129-6\_13} {\emph {\bibinfo {booktitle}
			{Artificial Intelligence and Intelligent Matter: Nanoscience, Soft Matter,
				Philosophy}}}\ (\bibinfo  {publisher} {Springer},\ \bibinfo {year} {2026})\
	pp.\ \bibinfo {pages} {257--271}\BibitemShut {NoStop}%
	\bibitem [{\citenamefont {Jeggle}\ and\ \citenamefont
		{Wittkowski}(2026)}]{jeggle2026intelligent}%
	\BibitemOpen
	\bibfield  {author} {\bibinfo {author} {\bibfnamefont {J.}~\bibnamefont
			{Jeggle}}\ and\ \bibinfo {author} {\bibfnamefont {R.}~\bibnamefont
			{Wittkowski}},\ }\bibfield  {title} {\bibinfo {title} {Intelligent matter
			consisting of active particles},\ }in\ \href
	{https://doi.org/10.1007/978-3-032-04129-6_14} {\emph
		{\bibinfo {booktitle} {Artificial Intelligence and Intelligent Matter:
				Nanoscience, Soft Matter, Philosophy}}}\ (\bibinfo  {publisher} {Springer},\
	\bibinfo {year} {2026})\ p.\ \bibinfo {pages} {273}\BibitemShut {NoStop}%
	\bibitem [{\citenamefont {Olsen}\ \emph {et~al.}(2026)\citenamefont {Olsen},
		\citenamefont {Tarama},\ and\ \citenamefont
		{L{\"o}wen}}]{olsen2026information}%
	\BibitemOpen
	\bibfield  {author} {\bibinfo {author} {\bibfnamefont {K.~S.}\ \bibnamefont
			{Olsen}}, \bibinfo {author} {\bibfnamefont {M.}~\bibnamefont {Tarama}},\ and\
		\bibinfo {author} {\bibfnamefont {H.}~\bibnamefont {L{\"o}wen}},\ }\bibfield
	{title} {\bibinfo {title} {Information bound on navigation speed in smart
			active matter},\ }\href {https://doi.org/10.1103/xs9l-1233} {\bibfield
		{journal} {\bibinfo  {journal} {Phys. Rev. Lett.}\ }\textbf {\bibinfo
			{volume} {137}},\ \bibinfo {pages} {157101} (\bibinfo {year}
		{2026})}\BibitemShut {NoStop}%
	\bibitem [{\citenamefont {Schwarzendahl}\ and\ \citenamefont
		{L\"owen}(2022)}]{Schwarzendahl/Lowen:2022}%
	\BibitemOpen
	\bibfield  {author} {\bibinfo {author} {\bibfnamefont {F.~J.}\ \bibnamefont
			{Schwarzendahl}}\ and\ \bibinfo {author} {\bibfnamefont {H.}~\bibnamefont
			{L\"owen}},\ }\bibfield  {title} {\bibinfo {title} {Anomalous cooling and
			overcooling of active colloids},\ }\href
	{https://doi.org/10.1103/physrevlett.129.138002} {\bibfield  {journal}
		{\bibinfo  {journal} {Phys. Rev. Lett.}\ }\textbf {\bibinfo {volume} {129}},\
		\bibinfo {pages} {138002} (\bibinfo {year} {2022})}\BibitemShut {NoStop}%
	\bibitem [{\citenamefont {Antonov}\ and\ \citenamefont
		{L\"owen}(2026)}]{Antonov/Lowen:2026}%
	\BibitemOpen
	\bibfield  {author} {\bibinfo {author} {\bibfnamefont {A.~P.}\ \bibnamefont
			{Antonov}}\ and\ \bibinfo {author} {\bibfnamefont {H.}~\bibnamefont
			{L\"owen}},\ }\bibfield  {title} {\bibinfo {title} {Temperature overshooting
			in the {M}pemba effect of frictional active matter},\ }\href
	{https://doi.org/10.1103/j4wb-1d9d} {\bibfield  {journal} {\bibinfo
			{journal} {Phys. Rev. E}\ }\textbf {\bibinfo {volume} {113}},\ \bibinfo
		{pages} {025407} (\bibinfo {year} {2026})}\BibitemShut {NoStop}%
	\bibitem [{\citenamefont {Debnath}\ \emph {et~al.}(2021)\citenamefont
		{Debnath}, \citenamefont {Chaudhury}, \citenamefont {Mukherjee},
		\citenamefont {Mondal},\ and\ \citenamefont {Ghosh}}]{debnath2021escape}%
	\BibitemOpen
	\bibfield  {author} {\bibinfo {author} {\bibfnamefont {T.}~\bibnamefont
			{Debnath}}, \bibinfo {author} {\bibfnamefont {P.}~\bibnamefont {Chaudhury}},
		\bibinfo {author} {\bibfnamefont {T.}~\bibnamefont {Mukherjee}}, \bibinfo
		{author} {\bibfnamefont {D.}~\bibnamefont {Mondal}},\ and\ \bibinfo {author}
		{\bibfnamefont {P.~K.}\ \bibnamefont {Ghosh}},\ }\bibfield  {title} {\bibinfo
		{title} {Escape kinetics of self-propelled particles from a circular
			cavity},\ }\href {https://doi.org/10.1063/5.0070842} {\bibfield  {journal}
		{\bibinfo  {journal} {J. Chem. Phys.}\ }\textbf {\bibinfo {volume} {155}},\
		\bibinfo {pages} {194102} (\bibinfo {year} {2021})}\BibitemShut {NoStop}%
	\bibitem [{\citenamefont {Upadhyaya}\ and\ \citenamefont
		{Akella}(2024)}]{upadhyaya2024narrow}%
	\BibitemOpen
	\bibfield  {author} {\bibinfo {author} {\bibfnamefont {A.}~\bibnamefont
			{Upadhyaya}}\ and\ \bibinfo {author} {\bibfnamefont {V.~S.}\ \bibnamefont
			{Akella}},\ }\bibfield  {title} {\bibinfo {title} {The narrow escape problem
			of a chiral active particle {(CAP):} an optimal scheme},\ }\href
	{https://doi.org/10.1039/d4sm00045e} {\bibfield  {journal} {\bibinfo
			{journal} {Soft Matter}\ }\textbf {\bibinfo {volume} {20}},\ \bibinfo {pages}
		{2280} (\bibinfo {year} {2024})}\BibitemShut {NoStop}%
	\bibitem [{\citenamefont {Baconnier}\ \emph {et~al.}(2022)\citenamefont
		{Baconnier}, \citenamefont {Shohat}, \citenamefont {L{\'o}pez}, \citenamefont
		{Coulais}, \citenamefont {D{\'e}mery}, \citenamefont {D{\"u}ring},\ and\
		\citenamefont {Dauchot}}]{baconnier2022selective}%
	\BibitemOpen
	\bibfield  {author} {\bibinfo {author} {\bibfnamefont {P.}~\bibnamefont
			{Baconnier}}, \bibinfo {author} {\bibfnamefont {D.}~\bibnamefont {Shohat}},
		\bibinfo {author} {\bibfnamefont {C.~H.}\ \bibnamefont {L{\'o}pez}}, \bibinfo
		{author} {\bibfnamefont {C.}~\bibnamefont {Coulais}}, \bibinfo {author}
		{\bibfnamefont {V.}~\bibnamefont {D{\'e}mery}}, \bibinfo {author}
		{\bibfnamefont {G.}~\bibnamefont {D{\"u}ring}},\ and\ \bibinfo {author}
		{\bibfnamefont {O.}~\bibnamefont {Dauchot}},\ }\bibfield  {title} {\bibinfo
		{title} {Selective and collective actuation in active solids},\ }\href
	{https://doi.org/10.1038/s41567-022-01704-x} {\bibfield  {journal} {\bibinfo
			{journal} {Nat. Phys.}\ }\textbf {\bibinfo {volume} {18}},\ \bibinfo {pages}
		{1234} (\bibinfo {year} {2022})}\BibitemShut {NoStop}%
	\bibitem [{\citenamefont {Chor}\ \emph {et~al.}(2023)\citenamefont {Chor},
		\citenamefont {Sohachi}, \citenamefont {Goerlich}, \citenamefont {Rosen},
		\citenamefont {Rahav},\ and\ \citenamefont {Roichman}}]{Chor/etal:2023}%
	\BibitemOpen
	\bibfield  {author} {\bibinfo {author} {\bibfnamefont {O.}~\bibnamefont
			{Chor}}, \bibinfo {author} {\bibfnamefont {A.}~\bibnamefont {Sohachi}},
		\bibinfo {author} {\bibfnamefont {R.}~\bibnamefont {Goerlich}}, \bibinfo
		{author} {\bibfnamefont {E.}~\bibnamefont {Rosen}}, \bibinfo {author}
		{\bibfnamefont {S.}~\bibnamefont {Rahav}},\ and\ \bibinfo {author}
		{\bibfnamefont {Y.}~\bibnamefont {Roichman}},\ }\bibfield  {title} {\bibinfo
		{title} {Many-body {S}zil\'ard engine with giant number fluctuations},\ }\href
	{https://doi.org/10.1103/PhysRevResearch.5.043193} {\bibfield  {journal}
		{\bibinfo  {journal} {Phys. Rev. Res.}\ }\textbf {\bibinfo {volume} {5}},\
		\bibinfo {pages} {043193} (\bibinfo {year} {2023})}\BibitemShut {NoStop}%
	\bibitem [{\citenamefont {Engbring}\ \emph {et~al.}(2023)\citenamefont
		{Engbring}, \citenamefont {Boriskovsky}, \citenamefont {Roichman},\ and\
		\citenamefont {Lindner}}]{engbring2023nonlinear}%
	\BibitemOpen
	\bibfield  {author} {\bibinfo {author} {\bibfnamefont {K.}~\bibnamefont
			{Engbring}}, \bibinfo {author} {\bibfnamefont {D.}~\bibnamefont
			{Boriskovsky}}, \bibinfo {author} {\bibfnamefont {Y.}~\bibnamefont
			{Roichman}},\ and\ \bibinfo {author} {\bibfnamefont {B.}~\bibnamefont
			{Lindner}},\ }\bibfield  {title} {\bibinfo {title} {A nonlinear
			fluctuation-dissipation test for {M}arkovian systems},\ }\href
	{https://doi.org/10.1103/physrevx.13.021034} {\bibfield  {journal} {\bibinfo
			{journal} {Phys. Rev. X}\ }\textbf {\bibinfo {volume} {13}},\ \bibinfo
		{pages} {021034} (\bibinfo {year} {2023})}\BibitemShut {NoStop}%
	\bibitem [{\citenamefont {Caprini}\ \emph {et~al.}(2024)\citenamefont
		{Caprini}, \citenamefont {Ldov}, \citenamefont {Gupta}, \citenamefont
		{Ellenberg}, \citenamefont {Wittmann}, \citenamefont {L{\"o}wen},\ and\
		\citenamefont {Scholz}}]{caprini2024emergent}%
	\BibitemOpen
	\bibfield  {author} {\bibinfo {author} {\bibfnamefont {L.}~\bibnamefont
			{Caprini}}, \bibinfo {author} {\bibfnamefont {A.}~\bibnamefont {Ldov}},
		\bibinfo {author} {\bibfnamefont {R.~K.}\ \bibnamefont {Gupta}}, \bibinfo
		{author} {\bibfnamefont {H.}~\bibnamefont {Ellenberg}}, \bibinfo {author}
		{\bibfnamefont {R.}~\bibnamefont {Wittmann}}, \bibinfo {author}
		{\bibfnamefont {H.}~\bibnamefont {L{\"o}wen}},\ and\ \bibinfo {author}
		{\bibfnamefont {C.}~\bibnamefont {Scholz}},\ }\bibfield  {title} {\bibinfo
		{title} {Emergent memory from tapping collisions in active granular matter},\
	}\href {https://doi.org/10.1038/s42005-024-01540-w} {\bibfield  {journal}
		{\bibinfo  {journal} {Commun. Phys.}\ }\textbf {\bibinfo {volume} {7}},\
		\bibinfo {pages} {52} (\bibinfo {year} {2024})}\BibitemShut {NoStop}%
	\bibitem [{\citenamefont {Antonov}\ \emph {et~al.}(2024)\citenamefont
		{Antonov}, \citenamefont {Caprini}, \citenamefont {Ldov}, \citenamefont
		{Scholz},\ and\ \citenamefont {L\"owen}}]{antonov2024inertial}%
	\BibitemOpen
	\bibfield  {author} {\bibinfo {author} {\bibfnamefont {A.~P.}\ \bibnamefont
			{Antonov}}, \bibinfo {author} {\bibfnamefont {L.}~\bibnamefont {Caprini}},
		\bibinfo {author} {\bibfnamefont {A.}~\bibnamefont {Ldov}}, \bibinfo {author}
		{\bibfnamefont {C.}~\bibnamefont {Scholz}},\ and\ \bibinfo {author}
		{\bibfnamefont {H.}~\bibnamefont {L\"owen}},\ }\bibfield  {title} {\bibinfo
		{title} {Inertial active matter with {C}oulomb friction},\ }\href
	{https://doi.org/10.1103/physrevlett.133.198301} {\bibfield  {journal}
		{\bibinfo  {journal} {Phys. Rev. Lett.}\ }\textbf {\bibinfo {volume} {133}},\
		\bibinfo {pages} {198301} (\bibinfo {year} {2024})}\BibitemShut {NoStop}%
	\bibitem [{\citenamefont {Antonov}\ \emph {et~al.}(2025)\citenamefont
		{Antonov}, \citenamefont {Musacchio}, \citenamefont {L{\"o}wen},\ and\
		\citenamefont {Caprini}}]{antonov2025self}%
	\BibitemOpen
	\bibfield  {author} {\bibinfo {author} {\bibfnamefont {A.~P.}\ \bibnamefont
			{Antonov}}, \bibinfo {author} {\bibfnamefont {M.}~\bibnamefont {Musacchio}},
		\bibinfo {author} {\bibfnamefont {H.}~\bibnamefont {L{\"o}wen}},\ and\
		\bibinfo {author} {\bibfnamefont {L.}~\bibnamefont {Caprini}},\ }\bibfield
	{title} {\bibinfo {title} {Self-sustained frictional cooling in active
			matter},\ }\href {https://doi.org/10.1038/s41467-025-62626-9} {\bibfield
		{journal} {\bibinfo  {journal} {Nat. Commun.}\ }\textbf {\bibinfo {volume}
			{16}},\ \bibinfo {pages} {7235} (\bibinfo {year} {2025})}\BibitemShut
	{NoStop}%
	\bibitem [{\citenamefont {Goerlich}\ \emph {et~al.}(2026)\citenamefont
		{Goerlich}, \citenamefont {Antonov}, \citenamefont {Olsen}, \citenamefont
		{Caprini}, \citenamefont {Scholz}, \citenamefont {L{\"o}wen},\ and\
		\citenamefont {Roichman}}]{goerlich2026particle}%
	\BibitemOpen
	\bibfield  {author} {\bibinfo {author} {\bibfnamefont {R.}~\bibnamefont
			{Goerlich}}, \bibinfo {author} {\bibfnamefont {A.~P.}\ \bibnamefont
			{Antonov}}, \bibinfo {author} {\bibfnamefont {K.~S.}\ \bibnamefont {Olsen}},
		\bibinfo {author} {\bibfnamefont {L.}~\bibnamefont {Caprini}}, \bibinfo
		{author} {\bibfnamefont {C.}~\bibnamefont {Scholz}}, \bibinfo {author}
		{\bibfnamefont {H.}~\bibnamefont {L{\"o}wen}},\ and\ \bibinfo {author}
		{\bibfnamefont {Y.}~\bibnamefont {Roichman}},\ }\bibfield  {title} {\bibinfo
		{title} {A particle-resolved rheological study of chirality transfer and odd
			transport},\ }\href {https://doi.org/10.48550/arXiv.2605.25136} {\bibfield
		{journal} {\bibinfo  {journal} {arXiv:2605.25136}\ } (\bibinfo {year}
		{2026})}\BibitemShut {NoStop}%
\end{thebibliography}
\end{document}